\documentclass[fleqn,usenatbib]{mnras}

\usepackage{newtxtext,newtxmath}

\usepackage[T1]{fontenc}
\usepackage{tablefootnote}

\DeclareRobustCommand{\VAN}[3]{#2}
\let\VANthebibliography\thebibliography
\def\thebibliography{\DeclareRobustCommand{\VAN}[3]{##3}\VANthebibliography}

\usepackage{graphicx}	
\usepackage{amsmath}	

\title[GECKOS Extraplanar Emission]{The GECKOS Survey: Extraplanar ionised gas in star-forming galaxies from eDIG to galaxy-scale winds}

\author[R. Elliott et al.]{R. Elliott$^{1,2}$, D.B. Fisher$^{1,2}$, B. Mazzilli Ciraulo$^{1,2}$, A. Fraser-McKelvie$^{3,2}$, M.~R. Hayden$^4$, M. Martig$^5$,
\newauthor J. van de Sande$^{6,2}$, A. J. Battisti$^{7,8,2}$, J. Bland-Hawthorn$^{9,2}$, A. D. Bolatto$^{10}$, T. H. Brown$^{11}$, B. Catinella$^{7,2}$,
\newauthor F. Combes$^{12,13}$, L. Cortese$^{7,2}$, T. A. Davis$^{14}$, E. Emsellem$^{3,15}$, D. A. Gadotti$^{16}$,
F. Pinna$^{20,21}$, T. H. Puzia$^{22}$,
\newauthor L. A. Silva-Lima$^{23}$, L. M. Valenzuela$^{24}$,
 G. van de Ven$^{25}$, 
\\
$^{1}$Centre for Astrophysics and Supercomputing, Swinburne University of Technology, Hawthorn, VIC 3122, Australia\\
$^{2}$ARC Centre of Excellence for All Sky Astrophysics in 3 Dimensions (ASTRO 3D), Australia\\
$^{3}$ European Southern Observatory, Karl-Schwarzschild-Stra{\ss}e 2, Garching, 85748, Germany \\
$^{4}$ Homer L. Dodge Department of Physics \& Astronomy, University of Oklahoma, 440 W. Brooks St., Norman, OK 73019, USA \\ 
$^{5}$ Astrophysics Research Institute, Liverpool John Moores University, 146 Brownlow Hill, Liverpool L3 5RF, UK \\
$^{6}$School of Physics, University of New South Wales, Sydney, NSW 2052, Australia \\
$^{7}$International Centre for Radio Astronomy Research (ICRAR), The University of Western Australia, M468, 35 Stirling Highway, Crawley, WA 6009, Australia \\
$^{8}$Research School of Astronomy and Astrophysics, Australian National University, Cotter Road, Weston Creek, ACT 2611, Australia \\
$^{9}$Sydney Institute for Astronomy, School of Physics, A28, The University of Sydney, NSW, 2006, Australia \\
$^{10}$Department of Astronomy, University of Maryland, College Park, MD 20742, USA \\
$^{11}$National Research Council of Canada, Herzberg Astronomy and Astrophysics Research Centre, 5071 W. Saanich Rd. Victoria, BC, V9E 2E7, Canada\\
$^{12}$Observatoire de Paris, LUX, CNRS, PSL University, Sorbonne University, 75014 Paris, France \\
$^{13}$ Coll\`ege de France, 11 Pl. Marcelin Berthelot, 75231 Paris, France \\
$^{14}$Cardiff Hub for Astrophysics Research \&\ Technology, School of Physics \&\ Astronomy, Cardiff University, Queens Buildings, Cardiff, CF24 3AA, UK \\
$^{15}$ Univ Lyon, Univ Lyon1, ENS de Lyon, CNRS, Centre de Recherche Astrophysique de Lyon UMR5574, F-69230 Saint-Genis-Laval France \\
$^{16}$ Centre for Extragalactic Astronomy, Department of Physics, Durham University, South Road, Durham DH1 3LE, UK \\
$^{17}$ Cosmic Dawn Center (DAWN), Denmark \\
$^{19}$ NSF's NOIRLab, 950 N. Cherry Avenue, Tucson, AZ 85719, USA \\
$^{20}$Instituto de Astrof\'isica de Canarias, calle Vía L\'actea s/n, E-38205 La Laguna, Tenerife, Spain\\
$^{21}$Departamento de Astrof\'isica, Universidad de La Laguna, Avenida Astrof\'isico Francisco S\'anchez s/n, E-38206 La Laguna, Spain \\
$^{22}$Instituto de Astrofísica, Pontificia Universidad Católica de Chile, Avenida Vicuña Mackenna 4860, 7820436, Macul, Santiago, Chile \\
$^{23}$Núcleo de Astrofísica, Universidade Cidade de São Paulo, Rua Galvão Bueno, 868, São Paulo, Brazil \\
$^{24}$Universitäts-Sternwarte, Fakultät für Physik, Ludwig-Maximilians-Universität München, Scheinerstr. 1, 81679 München,
Germany \\
$^{25}$ Department of Astrophysics, University of Vienna, T\"urkenschanzstra{\ss}e 17, 1180 Vienna, Austria \\
}

\date{Accepted XXX. Received YYY; in original form ZZZ}

\pubyear{\the\year{}}

\begin{document}
\label{firstpage}
\pagerange{\pageref{firstpage}--\pageref{lastpage}}
\maketitle

\begin{abstract}
We map the extraplanar gas, with $\sim$50-200~pc resolution, in nine star-forming galaxies using Multi-Unit Spectroscopic Explorer (MUSE) observations from the GECKOS VLT Large Program targeting edge-on galaxies with similar stellar mass as the Milky Way. The narrow range in stellar mass ($\pm0.35$~dex) of the GECKOS sample makes it ideal for studying trends with star formation rate (SFR). We find strong extraplanar emission reaching $\sim$2-8~kpc from the disk midplane in all targets with  $\rm{SFR}\geq$1~M$_{\odot}$~yr$^{-1}$.  Targets with SFR$\,\geq\,$5~M$_{\odot}$~yr$^{-1}$  have brighter, more extended H$\alpha$ emission compared to lower SFR targets.  In high-SFR systems, the gas velocity dispersion ($\sigma_{\rm H\alpha}$) shows a biconical morphology, consistent with the expectation of outflows. This agrees with previous works suggesting high velocity dispersion in a biconical shape is a good means to identify outflows. We find mixed results using line diagnostics ([OIII]$_{5007}$/H$\beta$ - [NII]/H$\alpha$ and $\sigma_{\rm H\alpha}$ - [SII]/H$\alpha$) to spatially resolve ionisation mechanisms across the extraplanar gas. The highest  [NII]/H$\alpha$ are the extraplanar gas of the highest SFR systems, yet main-sequence galaxies have the highest [OIII]/H$\beta$.  While the morphology of [NII]/H$\alpha$ may be useful to identify outflows, the absolute value of the line ratio alone may not  distinguish strong outflows from extraplanar gas of main-sequence galaxies. The ubiquitous extraplanar emission can be interpreted as the result of feedback, in the form of large-scale winds for starbursts or smaller-scale galactic fountains for main-sequence galaxies. Moreover, shock-heating may ionise gas at the interface of the disk and the circumgalactic medium, independent of the source of the gas.
\end{abstract}

\section{Introduction}

Extraplanar diffuse interstellar gas (eDIG) is commonly seen around star-forming galaxies \citep[e.g.][]{Dettmar1990,Rand1996,Rossa2003}.  Many mechanisms are invoked to explain its origin, including stellar feedback-driven galactic fountains \citep{Shapiro1976}, relic gas from galactic outflows \citep{LopezCoba2019}, leaky HII regions \citep{Haffner2009} and post-asymptotic giant branch stars in the thick stellar disk \citep{Flores2011}. Recent work on the EDGE-CALIFA survey \citep{Levy2019} argues that the vertical gas kinematics tend to favor an internal driving mechanism from the disk, i.e. stellar feedback, as the main source of eDIG. \cite{GonzalezDíaz2024} also find connections between the distribution of young massive star feedback and minor axis-parallel ionization profiles in a sample with modest SFR. 


Neither the observational differences between eDIG and galactic winds nor the strength of these differences is well established in main-sequence galaxies. The high velocity and turbulent internal motions of winds create a plausible environment for shocks, which may be diagnosed using optical emission line ratios \citep{Dopita1996}. Shock-excited gas typically exhibits elevated velocity dispersion and characteristic elevated optical emission line ratios. These signatures are used by studies to separate shock-ionized gas from star-formation-ionized gas. \citep{Rich_et_al._2010,Rich&Kewley&Dopita2011,Rich&Kewley&Dopita2015,Ho2014,Ho.et.al..2016,Kewley2019}. 

Galactic winds (i.e. outflows) are a specific type of extraplanar gas that play a key role in galaxy evolution. They are large-scale galaxy mass ejecta that have suppressive effects on star formation in the host galaxy. These ejecta are driven by supernovae explosions (SNe), radiation from young stars, and Active Galactic Nuclei (AGN). Outflows can carry interstellar medium (ISM) material far beyond the disk of the host galaxy \citep[reviews][]{Veilleux2020, Thompson.&.Heckman.(2024)}. Cosmological simulations find that mass-loss via outflows are a required component of the simulations, to match observations stellar masses and star formation rates (SFR) 
\citep{Somerville.&.Dave.(2015),Naab.&.Ostriker.(2017),Thompson.&.Heckman.(2024)}. There is, however, currently no theoretical consensus to describe the energetics of
outflows \citep[for example, see review by][]{Thompson.&.Heckman.(2024)}. Moreover, resolved surveys of galaxies with outflows are rare \citep[see discussion in][]{Veilleux.et.al..2020}. Recent work with optical IFU instruments is changing this \citep[e.g.][]{Reichardt2025}.

The last 15 years have seen a significant number of resolved observations of optical emission lines in extraplanar gas of both outflows and super-main-sequence systems  \citep{Sharp&BlandHawthorn2010,Bik2018,Lopez2020,ReichardtChu2022,ReichardtChu2024,McPherson2023,Watts2024}. Whilst these observations provide the opportunity to detect and study wind-shocks using line ratios, there is no clear consensus on the presence of shocks in the existing observations. \cite{Ho2014} argue that, based on observed line ratios, shock-ionized gas is very common in their sample of star-forming galaxies. Conversely, \cite{Bik2018} found, using MUSE imaging, that the gas above the center of the outflowing galaxy ESO338-IG04 has emission line ratios more consistent with photoionisation than shock-ionization. \cite{Chisholm2017} make a similar finding from using absorption lines to study a sample of outflowing galaxies. We note that moderately elevated emission line ratios in extraplanar gas do not {\em a priori} require shocks to exist: weak dust-shielding above galactic planes allow for harder radiation fields, resulting in elevated extraplanar line ratios \citep{Kewley2019}.

\begin{figure*}
\includegraphics[width=\textwidth]{
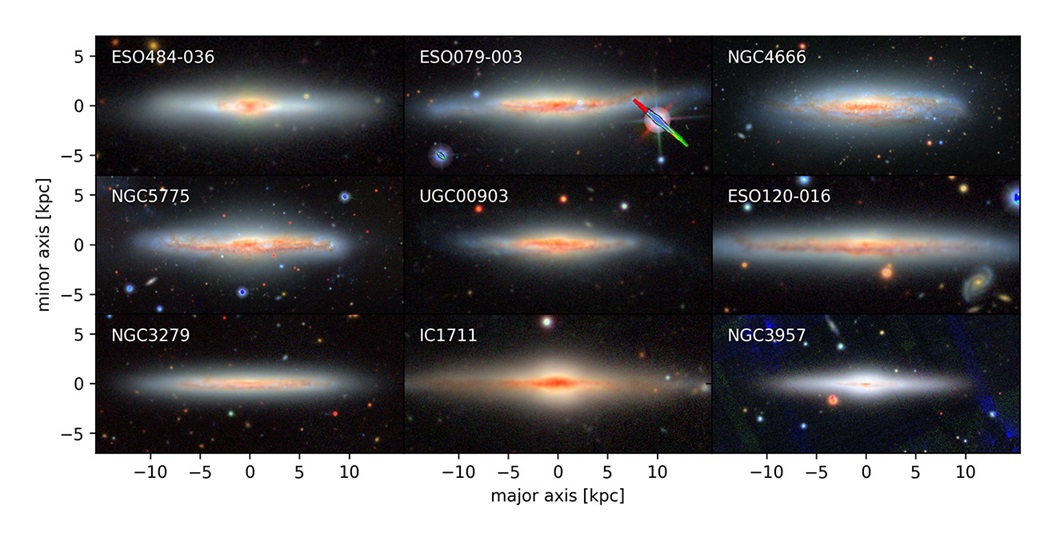} \caption{A collection of DESI Legacy Imaging Surveys (DR9) and Pan-STARRS-1 (DR2) colour images of the GECKOS galaxies covered in this paper. The images were produced using the $g$, $r$, and $z$ bands with the {\sc Astropy} package {\sc make\_lupton\_rgb} \citep{Lupton2004} and an asinh stretch. The galaxies were rotated to have the photometric major axis and the dust lane (and photometric major axis) oriented horizontally, and such that the dust lane (if offset) is to the lower side of the image.} \label{fig:DecALS}
\end{figure*}

This article uses GECKOS survey \citep[Generalising Edge-on galaxies and their Chemical bimodalities,  Kinematics, and Outflows out to Solar environments,][]{GECKOSpaper}  IFS data of nine edge-on galaxies
(Table \ref{tab:GECKOS_sample_table} + Fig.~\ref{fig:DecALS}). The galaxies have  masses within $\pm$0.3~dex of the Milky Way mass, $\sim5\times10^{10}$~M$_{\odot}$ \citep{BlandHawthorn_Gerhard_2016}, and SFRs varying over an order of magnitude 
 (SFR$\sim$0.2-8~M$_{\odot}$~yr$^{-1}$)
. We exploit deep MUSE observations from GECKOS to make resolved ($\sim50-200$~pc) studies of the kinematics and line-ratios of extraplanar gas across diverse star forming environments.

Sections \ref{sec:GECKOS}-\ref{sec:Observations} outline details of the GECKOS survey observations. Section~\ref{sec:data_management} describes the GECKOS data reduction process, and Section~\ref{sec:data_management_p2} describes the post-processing we performed on our reduced data. Sections \ref{sec:Ha Morphology}-\ref{sec:Ionization Morphology} present maps made from the reduced GECKOS data, highlighting kinematic and morphological differences between galaxies in our sample. Section~\ref{sec:region_definition} outlines how these maps can be systematically broken into different regions for analysis. Section~\ref{sec:Vertical Halpha surface brightness} compares surface brightness profiles from these regions in each galaxy, and \ref{sec:BPT line diagnostics} offers an emission line ratio analysis of our galaxy sample. \ref{sec:Discussion} offers a detailed discussion of the paper's results. Note that we have assumed a standard $\Lambda$CDM cosmology for this study (i.e. $H_0=68 km s^{-1}Mpc^{-1}$,$\Omega_m=0.3$).

\section{Methods} \label{sec:methods}

\subsection{GECKOS targets in this work} \label{sec:GECKOS}

The galaxies studied in this paper are part of the GECKOS survey \citep{GECKOSpaper}. The GECKOS survey is a 317~hr Large Program (PI: J. van de Sande) using the Multi-Unit Spectroscopic Explorer (hereafter MUSE) instrument on Unit Telescope 4 of the Very Large Telescope (VLT). The full GECKOS sample is 36 galaxies that are selected to have stellar mass within $\pm$0.3~dex of the Milky Way mass ($\sim5\times10^{10}$~M$_{\odot}$), are close to edge-on, at a distance of 10-70~Mpc. GECKOS targets are chosen to span a wide range in SFR, ($\sim2$~dex), making the sample ideal to search for mass-controlled correlations with SFR. The MUSE observations began in 2022 and concluded in 2025. This work studies the first nine star-forming galaxies, reduced and analysed as part of the GECKOS internal data release 1, as listed in Table~\ref{tab:GECKOS_sample_table}. 

The galaxy sample in this paper includes three galaxies that have previously been found to host galactic winds: NGC~4666 \citep{Dahlem1997}, NGC~5775 \citep{Heald2006},
and ESO~484-036 \citep{Veilleux_2005}. \cite{Heald2006} trace an outflow in NGC~5775 via a rotational lag between the disk and extraplanar regions. This lag is interpreted as a signature of outflowing material conserving momentum as it gradually expands into the halo. \cite{Heald2022} also reveal a mass outflow rate that is $\approx$40-80$\%$ of the SFR by modeling the galaxy's extended radio halo.
\cite{Dahlem1997} trace the NGC~4666 outflow by observing alignment between optical filaments, radio filaments, X-ray emission, and magnetic fields above the galaxy midplane. \citet{MazzilliCiraulo2025} carry out a detailed analysis of HI, CO and ionised gas for the NGC~4666 outflow using 
the same MUSE data used in this paper. They identify minor axis-parallel velocity profiles that are consistent with outflows. \cite{Dahlem1997} identify an outflow in ESO~484-036 via a characteristic X-shaped [NII]/H$\alpha$ emission line ratio enhancement emerging from the extraplanar emission in narrow-band imaging.

NGC~4666 and NGC~5775 are much closer than our other targets (around 15-20~Mpc compared to 25-70~Mpc). Therefore, MUSE tiles cover a smaller fraction of their extended emission , albeit with higher spatial resolution. We consider this difference when analysing the morphology of the extended emission.

\subsection{Observations}\label{sec:Observations}
Between 21/12/2022 and 21/8/2023, multiple observations were made for the nine galaxies in Table~\ref{tab:GECKOS_sample_table}. 
The exposure times were set to achieve a continuum signal-to-noise ($S/N$) above 40 at $\mu_g$ = 23.5~$\mathrm{mag~arcsec^{-2}}$, which is representative of the Solar environment of the Milky Way \citep{Melchior2007}. 

Most galaxies were observed using 1-3 pointings with a $1\times1$~arcmin$^2$ field-of-view, each covering different regions of the galaxy. We implemented tiling on nearby targets to reach a major-axis position with  $\mu_b$ = 23.5~mag~arcsec$^{-2}$. Each observation contained four $\sim$9-10~min object (O) observations and two $\sim$2-3~min offset sky observations (S), executed in an OSOOSO sequence. For targets with an SFR exceeding twice the main-sequence value at fixed stellar mass, we required MUSE tiles to cover a distance of at least 10~kpc from the galaxy midplane to identify extended winds. 

\begin{table*}

\begin{tabular}{lcccccccc} 
\hline
Galaxy & Distance & $\log(M_{\star})$ & SFR & $\Sigma_{SFR}$ & Spaxel Size & $R_{e}$\\
 & [Mpc]  & [M$_{\odot}$] & [M$_{\odot}$ yr$^{-1}$] & [10$^{-2}$ M$_{\odot}$ yr$^{-1}$   kpc$^{-2}$] & {[pc]} & {[kpc]} \\ \hline
ESO 484-036 & 69 & 10.5 & 8 & 10 & 200 & 5.1 \\
ESO 079-003 & 36 & 10.6 & 5 & 4 & 105 & 6.5 \\
NGC 4666 & 16 & 10.7 & 5 & 8 & 45 & 4.2 \\
NGC 5775 & 19 & 10.6 & 5 & 4 & 55 & 5.8 \\
UGC 00903 & 38 & 10.5 & 4 & 6 & 110 & 4.5 \\
ESO 120-016 & 50 & 10.6 & 3 & 1 & 146 & 8.7 \\
NGC 3279 & 30 & 10.3 & 1 & 1 & 87 & 5.5 \\
IC 1711 & 45 & 10.7 & 1 & 2 & 131 & 3.8 \\
NGC 3957 & 25 & 10.5 & 0.3 & 0.6 & 72 & 3.8 \\
\hline

\end{tabular}

\caption{Physical properties and MUSE observation parameters of the GECKOS galaxies analysed in this paper. Galaxies are ordered by decreasing SFR. Columns are as follows: galaxy names; distance \citet{Theureau2023} for ESO~079-003, and \citet{Tully2023} for all others; log(M$_{\star}$) is stellar mass from van de Sande {\em in prep}; SFR, derived from WISE Band-4 \citep{Cluver2017}; $\Sigma_{SFR}$ is SFR surface density, where $\Sigma_{SFR}=SFR/(\pi R_e^2)$; spaxel size is the resolution of the MUSE data for 0.6~arcsec resolution  data; and where $R_{e}$ is the effective radius (in arcsec) derived using Multi-Gaussian Expansion (MGE) models from \citet{Rutherford2025} in DECaLS r-band imaging \citep{Blum2016}. }
\label{tab:GECKOS_sample_table} 
\end{table*}

\subsection{Data Reduction} \label{sec:data_management} 
The data reduction process for GECKOS survey observations is described in \cite{FraserMcKelvie2025}. A full description of the GECKOS data reduction workflow will be presented in van de Sande et al. (in prep.). We use the python package \textsc{pymusepipe}\footnote{https://github.com/emsellem/pymusepipe} \citep{Emsellem2022} for data reduction, which is a wrapper for the MUSE Data Reduction Pipeline \citep[DRP,][]{weilbacher2020}. \textsc{pymusepipe} coordinated the multi-step data reduction process for observing block (OB), where each step is completed using the \textsc{esorex} package\footnote{https://www.eso.org/sci/software/cpl/esorex.html} \citep{esorex2015}.

The steps completed by \textsc{esorex} carries out bias and flat field corrections, and calibrates both wavelength and the line-spread function (LSF) for each raw pointing exposure. \textsc{esorex} then creates a geometry table to map the relative positions of each raw exposure pixel in the observed field of view. This mapping is followed by illumination corrections for each exposure,  based on the twilight flat exposures associated with the current OB, as well as the flatfield corrections made earlier. \textsc{esorex} performs
sky subtraction using offset sky exposures for the corresponding OB. The sky exposures are used to make a sky model, which is then fitted to each science frame. Three pointings of NGC\,4666 are archival data without sky frames. The sky subtraction procedure for NGC\,4666 involves using spaxels in object frames without strong emission lines. The above steps executed by \textsc{esorex} result in a datacube for each exposure. This data reduction process is described in more detail by \citet{MazzilliCiraulo2025}. 

We use \textsc{pymusepipe} and \textsc{spacepylot} to align the world coordinate system (WCS) of each exposure. Following an automated alignment routine, we corrected remaining spatial offsets between  galaxy exposures by aligning a corresponding r~band image from 
either the DESI Legacy image survey \citep{Dey2019} or the Pan-STARRS survey \citep{Chambers2016}. After alignment,  We stack each exposure datacube using MPDAF \citep{Bacon2016} to make a complete datacube for each each OB. We then mosaic these complete OB cubes together to make a complete datacube for one galaxy. 

\subsection{Emission Line Measurement} \label{sec:data_management_p2}

Emission line measurement and continuum subtraction follows the same procedure as used in \cite{MazzilliCiraulo2025}. Continuum subtraction was performed for each galaxy datacube using the {\sc nGIST} pipeline \citep{FraserMcKelvie2025}, which is an extension of the original GIST pipeline \citep{Bittner2019}.{\sc nGIST} builds on a number of works, which build processes compiled into the software \citep{Cappellari2003,Cappellari2004,Kuntschner2006,Vazdekis2010,Cappellari2017,Cappellari2023}. {\sc nGIST} is a modular pipeline for the analysis of integral field spectroscopic galaxy data. In this work, we employ the Continuum module, which creates continuum-only and line-only cubes after fitting the spectrum with a set of SSPs and masking emission lines. 
We use the \cite{Walcher2009} model for continuum fitting and subtraction, as this produces the smallest residuals in H$\beta$ absorption compared to other models we tested \citep{Vazdekis1999,Bruzual_and_Charlot_2003}. 
Following testing in previous work on resolved outflows \citep{ReichardtChu2022,McPherson2023}, we Voronoi bin to continuum $S/N=7$, and fitted the rest-frame wavelength range 4750-7100\AA.  The Milky Way extinction is removed using the \cite{Cardelli1989} extinction law. The continuum model fitted to each Voronoi bin is then subtracted from the corresponding spaxels at the 
native MUSE resolution of 0.2". In all targets there is only sufficient continuum S/N to use this method near to the galaxy
$z\lesssim \pm$1.5-2~kpc. Beyond this minor axis height there is no discernible shape to the continuum. In the Gaussian fitting  of the emission lines, described below, we include a constant flux offset for all fits. This handles imperfections in continuum removal, and those spaxels in which the S/N is too low to estimate continuum using pPXF.  

Continuum removal at wavelengths near and above $\sim$6700~\AA\ is complicated by several sky features. These features become increasingly common for larger wavelength. This then means that sky subtraction generates increasingly large residuals and imperfections redward of $\sim$6700~\AA .  These sky features alter the continuum from the stellar population modeling. Visual inspection of the pPXF model compared to the observed continuum shows poor representation of the pPXF fit to continuum. This is likely more noticeable in the GECKOS program, as we target low surface brightness emission lines at larger scale-height, where sky lines may have greater impact. We, therefore, follow a different continuum removal approach for different emission lines. For H$\beta$ and H$\alpha$ we must use the pPXF models to remove absorption. Moreover, the sky subtraction is well behaved in the these wavelengths and does not have an impact that we measure on the continuum model. Using the same pPXF model to remove continuum below [SII]~$\lambda \lambda$6716,31 introduces a significant systematic uncertainty. We therefore apply a local correction for continuum under the [SII]~$\lambda \lambda$6716,31 emission lines. 

As described in \cite{MazzilliCiraulo2025}, we identify a  bandpass of $\sim$50~\AA\ around the [SII]~$\lambda \lambda$6716,31 emission lines in which to estimate a local continuum correction. The bandpass of $\sim$50~\AA\ is chosen to be small enough such that the emission line is well approximated by a simple function (linear or constant). We find that in all targets the slope of the spectrum across this area is small compared to the noise in the spectrum, and is thus consistent with negligible gradient. We remove the continuum locally by a simple linear fit to the continuum. This results in a flat continuum, with no measurable gradient, around the [SII]~$\lambda \lambda$6716,31 emission lines. We tested using a constant value instead of a straight line, and in our tests these return similar fluxes as the linear continuum removal. We test the impact of this method by measuring the change in continuum over level in a 200~\AA\ region surrounding [SII] in spaxels both in the galaxy and in the outflow. We then compare this to the noise from the same region. We find that in the disk the continuum varies by $\sim$20\% of the scatter in the continuum, and in the outflow regions this reduces to $\sim$10\%. This small difference reflects the fact that the continuum near to [SII]~$\lambda \lambda$6716,31 emission lines does not vary significantly across the small wavelength range surrounding these emission lines. Removing the continuum with this method, therefore, does not introduce measurable sources of uncertainty, and generates fewer systematic uncertanties than using pPXF for the [SII] continuum. Target ESO~484-036, the most distant in this sample, contains a strong sky-line that heavily contaminates the [SII] doublet, preventing [SII] analysis for this target.

We also reiterate that for all emission lines at larger distance from the galaxy, $z\gtrsim \pm$1.5-2~kpc, there is insufficient continuum S/N to use any formalized fitting technique. This difference is, therefore, moot for most of the extraplanar emission. The Gaussian fits include a constant offset in the fit, which is sufficient to correct continuum in these more distant regions.

After the continuum removal, each galaxy datacube is rotated such that the galaxy major axis is horizontal in the data cube. In addition, cubes are spatially rebinned by combining pixels in 3$\times$3 squares. This rebinning yields a 0.6~arcsec resolution, which is a typical seeing value at the the VLT.

We performed single Gaussian fits to emission lines in the 3$\times$3 rebinned datacubes using the Python package {\sc threadcount}\footnote{https://threadcount.readthedocs.io}, described in previous work \citep{McPherson2023,Hamel2024,Reichardt2025,MazzilliCiraulo2025}. We take several steps to optimise for speed. The software uses a bespoke  version of the \texttt{nelder} minimisation algorithm, and a bespoke \footnote{This version of {\sc threadcount} remains in development, but is available on request from D.B.~Fisher: dfisher@swin.edu.au}version of the Python package \texttt{lmfit}. 

We run 10 iterations in which the spectrum is varied and refit. We make these variations according to a normal distribution, whose standard deviation is set by the observed variance. We, however, only employ MC iterations on low S/N regions of the galaxy. For spaxels with $S/N>30$, we find that there is no advantage to MC iterations.  At high S/N, fits with MC iterations return very similar parameters and uncertainty estimates as fits that do not employ the MC iteration (as single minimization run). For those spaxels with $S/N<30$, the fit parameters and uncertainty are then taken as the mean and standard deviation of those fits. 

We fitted the H$\alpha$, H$\beta$, $\mathrm{[OIII]}\lambda 5007$, $\mathrm{[NII]}\lambda 6548$, $\mathrm{[NII]}\lambda 6583$, $\mathrm{[SII]}\lambda 6716$, and $\mathrm{[SII]}\lambda 6731$ lines for each data cube. For the results presented in this paper, we only considered emission lines in which the flux is $3\times$ greater than the fit uncertainty. We applied the same fitting procedure to versions of the data cubes that were binned to a spatial resolution of 500~pc. Both the 3x3 bin and the 500~pc bin resolution cubes will be used in this paper. 

We apply extinction corrections to each fitted spaxel, by creating $A_V$ maps from spaxel-by-spaxel Balmer decrement measurements and \cite{Cardelli1989} extinction curve. We only used spaxels with fitted H$\alpha$ and H$\beta$ fluxes greater than $5\times$ the flux uncertainty to create $A_V$ maps. Spaxels with insufficient S/N for an extinction estimate were interpolated from nearby values. The interpolation may reduce the true variation of line fluxes in the data. These $A_V$ maps were then used to correct emission line fluxes in each spaxel. 

Velocity dispersions in all figures are shown with the instrumental dispersion subtracted in quaderature from the fitted disperson to each spaxel, where $\sigma_{inst}\sim49$~km~s$^{-1}$. 


\section{Emission Line Maps}
\label{sec:Mapping Properties of the GECKOS Galaxy Sample}

\begin{figure*}
    \includegraphics[width=0.87\textwidth]{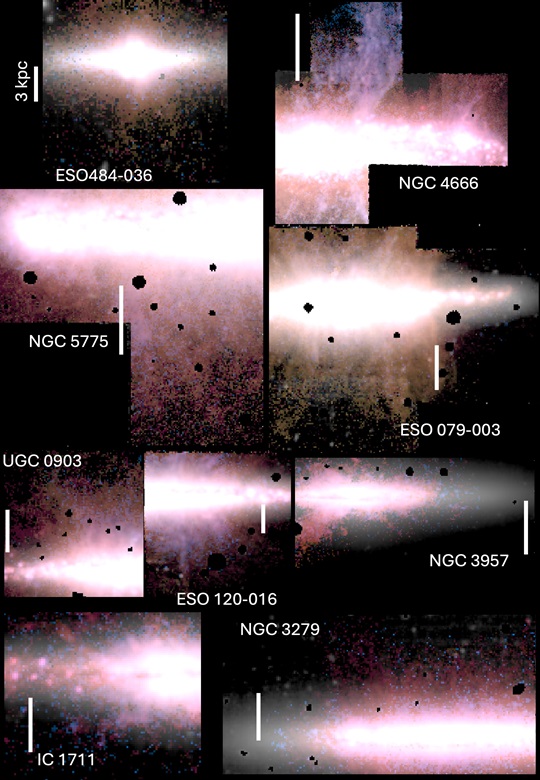}
    \caption{Four colour emission line plus continuum images of GECKOS targets studied in this paper. In all figures, H$\alpha$ emission is shown in red, [NII]~6583 in yellow, [OIII]~5007 in blue, and R~band continuum in white. The white line in each panel represents a $\sim$3~kpc scale. Extraplanar emission is visible in all targets, often forming multi-kiloparsec filaments extending out from the disk.} \label{fig:4color}
\end{figure*}

Fig.~\ref{fig:4color} shows the 4 color maps of targets in this work. The white, diffuse light represents the $\sim$800~nm continuum emission generated from MUSE cube. Emission lines are OIII~5007 (blue), H$\alpha$ (red) and [NII]~6583 (yellow). For display purposes only, we use a very low S/N cut of $2\sigma$ in Fig.~\ref{fig:4color}, to err on showing all emission. For all targets, filaments of gas extend upward away from the mid-plane of the $\sim$800~nm continuum image, typically this goes beyond 3~kpc from the galaxy. Similar to previous works, this figure makes it clear that extraplanar ionised gas emission is a common property of all star forming galaxies in our sample \citep[e.g.][]{Rossa2003,Levy2019}. The panels of the figure are arranged by SFR, highest at the top-left and lowest at the bottom-right. This makes a clear correlation of the extension of the extraplanar gas with galaxy SFR easy to see. 

The morphology of NGC~4666 is particularly interesting \citep[for a more in depth description see][]{MazzilliCiraulo2025}. It is the nearest galaxy in the sample, which allows for the detailed filaments to be resolved. Individual streams of gas with width similar to the resolution limit ($\sim$45~pc) are visible, they extend roughly 6~kpc from the galaxy. This may be similar to the plume morphologies seen in PAH emission from JWST observations \citep{Fisher2024}. Similar plume-like structures are visible in ESO~079-003 and even ESO~120-016, the latter of which has a relatively low SFR
($\sim$3~M$_{\odot}$~yr$^{-1}$.

We note that NGC~5775 is a similar distance to NGC~4666 and has a more smooth distribution of extraplanar emission. This galaxy has a well-studied outflow \citep[e.g.][]{Heald2006}. It has been shown to have radio continuum consistent with emission from a significant population of cosmic rays \citep{Heald2022}, which could act to smooth outflow substructure.  \cite{Stein2019} argues for the presence of cosmic rays in the halo of NGC~4666, but the relative impact of cosmic rays between NGC~4666 compared to NGC~5775 is uncertain. It remains unclear what sets the differences in substructure of outflows. Recent work on cloud-scale structures made with JWST, has been shown to direclty test outflow theories \citep{Fisher2024}. More work, with higher spatial resolution than this data, would be informative.

\begin{figure*}
\includegraphics[width=\textwidth]{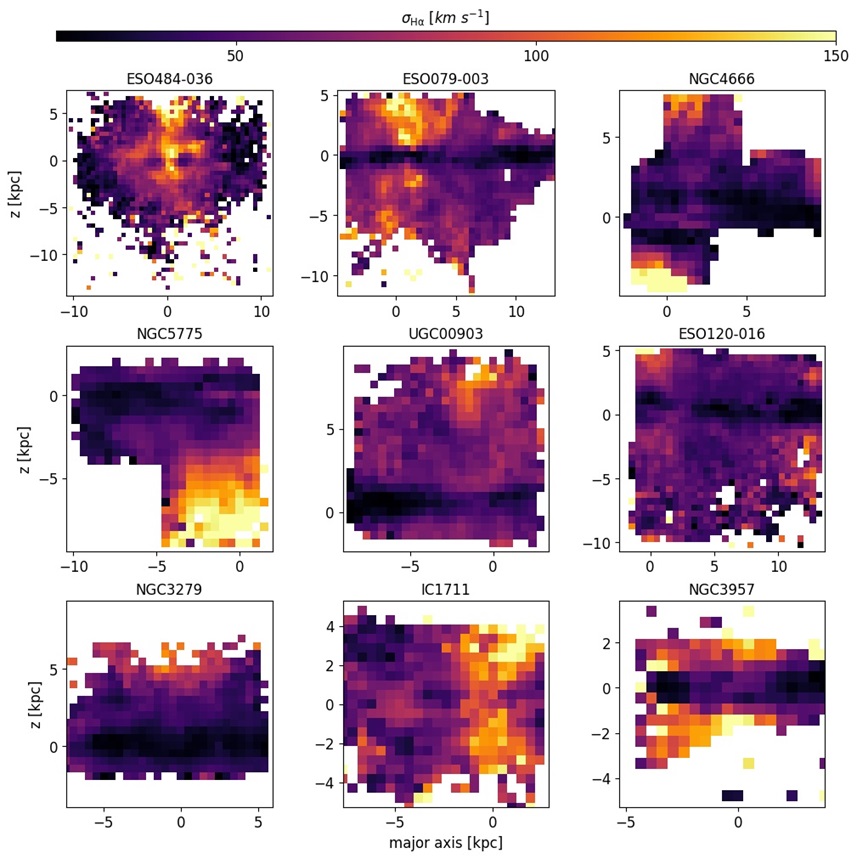} 
\caption{H$\alpha$ velocity
dispersion maps made from 500~pc resolution data for each galaxy.  The points where the vertical and horizontal axes =0 mark the approximate location of the galactic midplane and rotational axis respectively.}
\label{fig:maps_500pc}
\end{figure*}

Fig.~\ref{fig:maps} displays 0.6~arcsec resolution maps of H$\alpha$ flux, H$\alpha$ velocity dispersion, [NII]/H$\alpha$, and H$\alpha$ line-of-sight velocity for each galaxy in our sample. Fig.~\ref{fig:maps_500pc} displays alternative velocity dispersion maps ($\sigma_{H\alpha}$), in which the MUSE data for each galaxy has been binned to $\sim$500~pc per spaxel and then refit with {\sc threadcount}. The first binning scheme, 3$\times$3, simply restructures the cubes to match seeing, while the coarser binning, to 500~pc, is intended to increase $S/N$ of line measurements and probe the lowest surface brightness more robustly. In all maps, we observe H$\alpha$ flux decreasing with height from the midplane, whilst H$\alpha$ velocity dispersion and [NII]/H$\alpha$ generally increase. 

\subsection{H$\alpha$ Morphology}\label{sec:Ha Morphology}

Fig.~\ref{fig:maps} shows we observe H$\alpha$ emission at heights greater than $\pm$4~kpc in all but one target, NGC~3957. This galaxy possesses both the lowest SFR and lowest extraplanar H$\alpha$ surface brightness in our sample. All targets with SFR$\geq$1~M$_{\odot}$~yr$^{-1}$ have extended H$\alpha$ emission. We will discuss correlations of the extended emission with galaxy SFR in later sections of this paper. 


\begin{figure*}
\includegraphics[width=\textwidth]{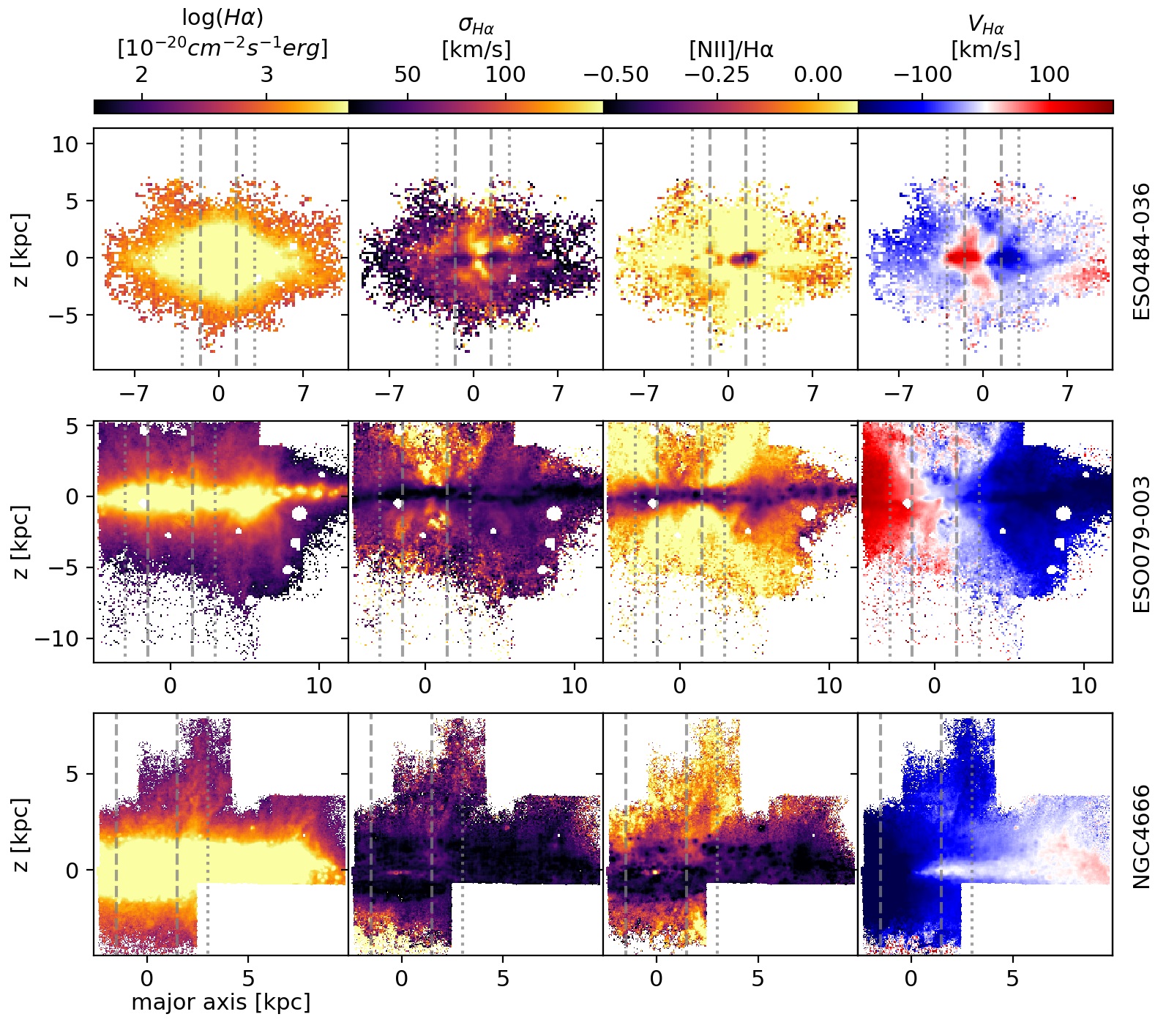} 
\caption{Maps of
log(H$\alpha$) flux (left), H$\alpha$ velocity
dispersion that has been corrected for instrumental dispersion (middle-left), and log([NII]/H$\alpha$)
flux ratio (middle-right) and velocity (right) made for each galaxy in our sample. Note that the
galaxies in this figure are ordered by decreasing SFR. Spatial axes are in units of kpc. The points 
where the vertical and horizontal axes equal 0 kpc mark the approximate location of the galactic midplane and rotational center respectively. We mark the 1.5~kpc and 3.5~kpc points on the major axis with vertical gray dashed and dotted lines respectively.}
\label{fig:maps}
\end{figure*}

\begin{figure*}
\includegraphics[width=\textwidth]{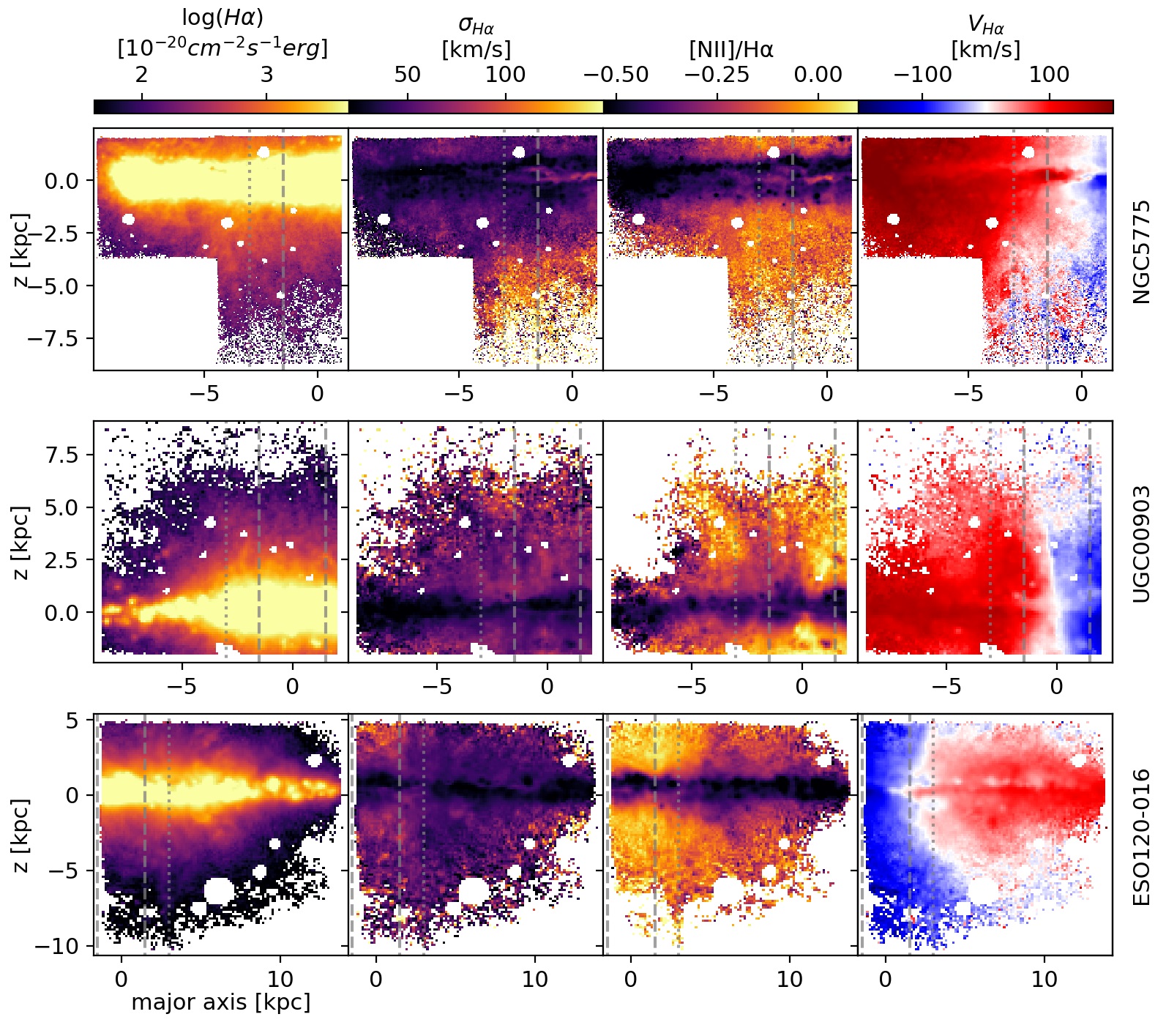}
\contcaption{}
\label{fig:maps2}
\end{figure*}

\begin{figure*}
\includegraphics[width=\textwidth]{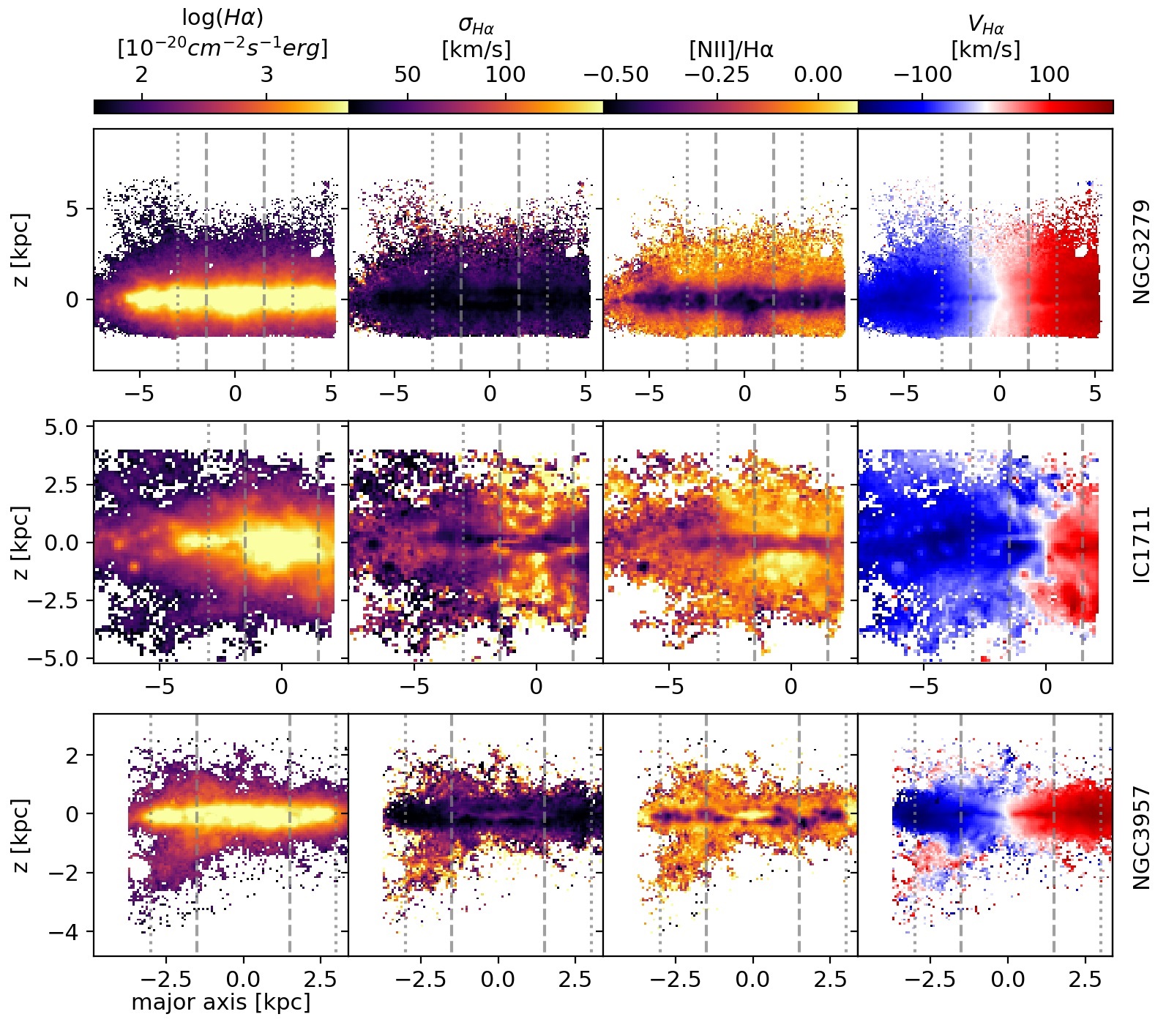} 
\contcaption{}
\label{fig:maps3}
\end{figure*}

Many galaxies in our sample show filamentary substructures in the H$\alpha$ flux, which extend at near-to perpendicular angles from the disk midplane outward into the extraplanar regions. We see these features in ESO~484-036, NGC~4666, NGC~5775, ESO~079-003, IC~1711, UGC~00903, ESO~120-016. Comparable features in NGC~3279 do not extend as far. Such substructure has likewise been discussed in M~82 using JWST \citep{Fisher2024} and HST~H$\alpha$ imaging \citep{Lopez2025}. This suggests similarity of GECKOS outflows to M~82. 

NGC~4666 is described in \cite{MazzilliCiraulo2025} to show the traditional biconical structure, which the MUSE FOV here shows part. Similarly NGC~5775 has been shown to have a biconical wind \citep{Heald2006,Heald2022}, of which a portion is covered. ESO~484-036 is shown in \cite{2002ApJ...565L..63V} to likewise show biconical structure, which is visible in the figure. We report that ESO~079-0003 shows biconical filaments in H$\alpha$ flux that are consistent with the traditional bicone morphology. 

NGC~3957, the lowest SFR galaxy in our sample (0.3~M$_{\odot}$~yr$^{-1}$), has an off-center filament of gas extending over $\sim$2~kpc from the disk. This is dissimilar to the extraplanar emission in other GECKOS targets, and that observed in low SFR systems with MUSE \citep[e.g.][]{GonzalezDíaz2024}. Like all other targets in this sample, the H$\alpha$ surface brightness of NGC~3957 peaks at the galactic center. We check that WISE W3 shows a similar central peak, which gives an SFR indicator that is more robust to extinction. 
The off-center filament in NGC~3957 is not connecting to the strongest star formation in the galaxy, but instead to a relatively lower SFR region of the disk.  Moreover, the kinematics of this filament are irregular. As shown in the rightmost panel of Fig.~\ref{fig:maps3}, the gas at $-$1~kpc below the midplane is counter-rotating with respect to the disk at the same major-axis position. This feature sets this off-center plume apart from other filaments in our sample. 

\subsection{Velocity Dispersion Maps}
In all targets the velocity dispersion increases at greater vertical distance from the midplane. In several galaxies (e.g. ESO~484-36 and ESO~079-003) the elevated velocity dispersion has a biconical shape over the center of the galaxy. 
For this analysis, we include discussion of the 500~pc resolution dispersion maps in Fig.~\ref{fig:maps_500pc}. The 500~pc maps allow us to probe further from the galaxy center, albeit at a $\sim2.5-10\times$ coarser resolution than the 0.6~arcsec resolution maps in Fig.~\ref{fig:maps}. We remind the reader that the instrumental dispersion is subtracted in quaderature from the fitted disperson. 

We also note for physical interpretation of $\sigma_{H\alpha}$ that in extraplanar gas the velocity dispersion may not indicate turbulence, especially at MUSE spectral resolution. \cite{McPherson2023} describe how the line-of-sight expansion of a bicone when convolved with relatively low spectral resolution (R$\lesssim 3000$) could appear as a broadened line rather that line-splitting. We will consider this in the text when appropriate.


We first consider galaxies with higher SFR$\gtrsim 5 $~M$_{\odot}$~yr$^{-1}$.  Galaxies ESO~484-036, NGC~5775 and ESO79-003 show extensive extraplanar regions with $\sigma_{\rm H\alpha} > 100$~km~s$^{-1}$ that is above $z\gtrsim$1~kpc and within 1~kpc of the major-axis center.  In ESO~079-003, the high dispersion regions form a triangular structure that extends up to $\sim$10~kpc from the galaxy midplane. This structure is visible in both the 0.6~arcsec resolution maps (Fig.~\ref{fig:maps}) and 500~pc resolution maps (Fig.~\ref{fig:maps_500pc}). At a constant height of $z\sim$1~kpc, the velocity dispersion drops to values of $\sim$40-60~km~s$^{-1}$ outside this triangular region. A similar triangular region may be present in ESO~484-036, as can be seen most clearly in the 500~pc maps (Fig.~\ref{fig:maps_500pc}). We cannot fully describe the morphology for NGC~5775; this is because the extraplanar MUSE pointings do not cover the entire outflow, and only extend $\sim$5~kpc from the galaxy center (on the major-axis).  We do, however, identify high dispersion gas ($\sigma_{\rm H\alpha}>100$~km~s$^{-1}$) above the galactic center of NGC~5775, which declines near the edge of the field-of-view at $\sim$4~kpc from the major axis center. NGC~4666 shows a strong rise in $\sigma_{\rm H\alpha}$ on the lower side of the 500~pc map, but limited spatial coverage obscures the shape of this $\sigma_{\rm H\alpha}$ enhancement. Likewise on the top side of NGC~4666, the coverage barely includes the center of the major-axis in the wind region, which means that detecting a centrally located rise in $\sigma_{\rm H\alpha}$ is difficult. Overall, in the high SFR super-main sequence systems with sufficient spatial coverage, there is evidence of a biconical high dispersion structure in the 500~pc resolution velocity maps.

We now consider the lower SFR galaxies. UGC~00903 and ESO~120-016 also have elevated dispersions in extrplanar gas above the disk center (on the major-axis), which have peak values of $\sigma_{\rm H\alpha}\sim$70~km~s$^{-1}$. Higher SFR systems commonly reach $\sigma_{\rm H\alpha}\sim$100-150~km~s$^{-1}$, which can be twice the highest values in most of the low SFR systems. Fig.~\ref{fig:maps_500pc} also displays $\sigma_{\rm H\alpha} > 100$~km~s$^{-1}$ regions above the planes of NGC~3957 and NGC~3279, but these do not share the triangular shapes seen in galaxies like ESO~079-003 and ESO~484-036.

IC~1711 displays a high dispersion triangular region. Unlike other galaxies with this feature, IC~1711 has a modest SFR ($\sim1$~M$_{\odot}$~yr$^{-1}$) and little H$\alpha$ emission beyond a minor axis height $\sim$4~kpc. One interpretation is that IC~1711 could contain historic wind activity. Simulations show this could leave a mark on the extraplanar gas (see \citealp{Pillepich2021}). This activity could be from alternate possibilities, such as a weak AGN that is currently driving a low mass outflow. We discuss the case of IC~1711 in more detail later.  

\cite{McPherson2023} discuss the use of velocity dispersion in extraplanar emission as a diagnostic for galactic winds, separating it from extraplanar emission not associated with the wind. Similar conclusions can be drawn from inspection of outflows discussed in \cite{Bik2018} and \cite{Watts2024}. \cite{McPherson2023} argue that high dispersion signatures are generated by the expansion of a biconical structure in ionised gas, which, when observed with modest spectral resolution (R$\sim2000-3000$), appears as a broad Gaussian emission line. Turbulence generated within the outflow would further increase this velocity dispersion. Our velocity dispersion results are consistent with this, albeit with some caveats. ESO~484-036, NGC~4666, NGC~5775 and ESO~079-003 all have SFR$\geq 5$~M$_{\odot}$~yr$^{-1}$, and the first 3 of these galaxies have known outflows \citep{Veilleux_2005,Heald2006}. NGC~5775 and NGC~4666 have limited MUSE coverage of the wind, making it difficult to identify any biconical dispersion structures. However, the 500~pc image does show high dispersion values on the lower side of NGC~4666, as well as increased dispersion around $z\sim 5-7$~kpc on the top side. We see likewise for $z$<-4 kpc in NGC~5775. These high dispersion values may trace some triangular dispersion structures, but we cannot be certain without a wider field-of-view.

These first results from GECKOS galactic winds, especially in Fig.~\ref{fig:maps_500pc}, allow us to create more formalized criteria for outflows. We rely on the assumption that at the same stellar mass, galaxies with higher SFR are expected to be more likely to have strong outflows \citep[e.g.][]{Heckman_et_al._2015,Chisholm_et_al._2015,Veilleux.et.al..2020,Thompson.&.Heckman.(2024)}. More specifically in our sample, the literature has described our high SFR systems as having outflows (see references above). We then interpret that our galaxies with SFR$\geq$5~M$_{\odot}$~yr$^{-1}$ are outflow systems. Under this assumption we find the following:  

\noindent (1) {\bf Elevated Velocity Dispersion as an Outflow Identifier}: We find that  when the velocity dispersion of the outflow is above $\sim100$~km~s$^{-1}$ this system is most likely a stronger, biconical outflow. We note that, as seen in other works, we observe at least some increase in velocity dispersion with increasing distance from the midplane for all galaxies in our sample, but the outflow systems have a larger increase. 

\noindent (2) {\bf Velocity Dispersion Morphology as an Outflow Identifier}: We find that there is a triangular/biconical morphology in $\sigma_{\rm H\alpha}$ above most high-SFR galaxies.  Our results are therefore consistent with previous arguments that biconical morphology in a velocity dispersion map appears to be a good indicator that this gas is associated to a strong biconical outflow \citep[e.g.][]{McPherson2023}. We add the caveat that
some galaxies with such morphology, as found in IC~1711, are not fully understood and may have
an alternative origin.

\subsection{Ionization Morphology}
\label{sec:Ionization Morphology}

The [NII]/H$\alpha$ maps, shown in the 3$^{rd}$ column of Fig.~\ref{fig:maps}, of each galaxy allow us to visualise the morphology of the ionisation in the extraplanar gas. The classic `X-shape' structure in this emission line ratio, where arms of the X extend outward from the galaxy centre, has been described as indicative outflows for many years \citep[e.g.][]{Veilleux_2005}.  This is most clear in ESO~079-003, which shows higher levels of [NII]/H$\alpha$ in the limbs of the outflow. Though we note that in this target the bright H$\alpha$ knot located at roughly $\sim$6-7~kpc on the major-axis has a low [NII]/H$\alpha$, and is not part of the high-ionisation X-shape. In ESO~484-036, ESO~079-003, IC~1711, UGC~00903 and ESO~120-016, [NII]/H$\alpha$ peaks above the galaxy center (seen within major axis cuts of 1-2 kpc). As with velocity dispersion, it is difficult to interpret [NII]/H$\alpha$ in NGC~4666 and NGC~5775 due to the limited field-of-view coverage. We note that the morphologies of [NII]/H$\alpha$ of two of the lower SFR targets (NGC~3279 and NGC~3957) differ strongly from higher SFR galaxies like ESO~079-003. 

We note again the difference of IC~1711 to other low-SFR systems. IC~1711 has the second lowest SFR in our target list, yet the morphology of the [NII]/H$\alpha$ ratio more suggests a high ionization cone above the galaxy center. This again suggests the possibility for an alternate source of the extraplanar emission, as discussed above and will be discussed below in the Discussion.

\subsection{Regions of Extraplanar and Disk Emission}
\label{sec:region_definition}

\begin{figure}
\includegraphics[width=\columnwidth]{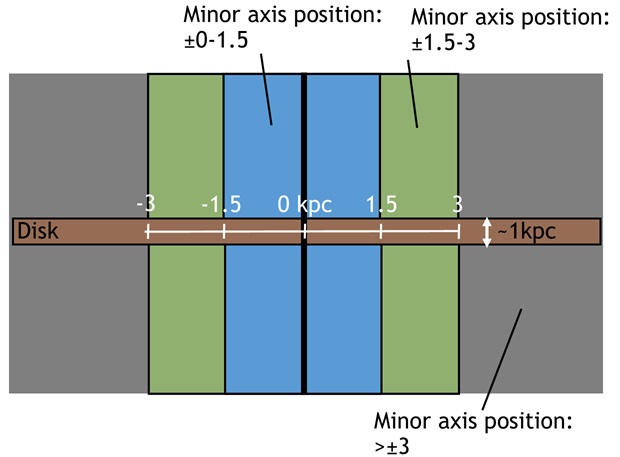}
\caption{Illustration of the regions defined for each galaxy in this
paper. Brown highlights the disk region, blue highlights the "major axis: $\pm$0-1.5~kpc" region, green highlights the "major axis: $\pm$1.5-3~kpc" region, and the grey area outside of these highlights is the "major axis: $>\pm3$ kpc" region. When we project these regions onto an observation of an edge-on galaxy, the "disk" region aligns with the galactic midplane, and the "0 kpc" mark aligns with the galactic center.}
\label{fig:regions}
\end{figure}

For our analysis of the ionised gas emission line ratios, we divide each galaxy into four empirically selected regions. These regions are motivated by previous observations of outflows, where hotter, more ionized gas is expected over the galaxy center and cooler/lower ionisation gas is found at further distances along the major axis \citep{Shopbell_1998,Veilleux_2005,Lopez2020,Veilleux.et.al..2020}. Our sample, which includes a range in SFR and likely a range in galactic wind strength, can emprically determine if this structure is connected to the presence of stronger winds.  We illustrate the regional seperations, described below, in Figure \ref{fig:regions}. The {\em disk} region is designed to align with the midplane of the galaxy, and capture ionization in the disk. It is not meant to include extraplanar emission. We define a 1~kpc thick region that extends along the full major-axis of the galaxy. This thickness is similar to the size of regions described as disk in previous outflow studies \citep[e.g.][]{McPherson2023}.  The extraplanar emission is divided into 3 regions based on the distance to the major-axis galaxy center. The boundaries for these separations based on {\em ad hoc} inspection of Fig.~\ref{fig:maps}, with special emphasis on the velocity dispersion. Based on what we have already discussed, we therefore expect a decrease in [NII]/H$\alpha$ and $\sigma_{\rm H\alpha}$ moving from the regions above the galaxy center towards those horizontally offset from the prior region. These definitions are intended to help us better quantify this decrease.




\subsection{Vertical H$\alpha$ surface brightness profiles}\label{sec:Vertical Halpha surface brightness}

\begin{figure*}
\includegraphics[width=\textwidth]{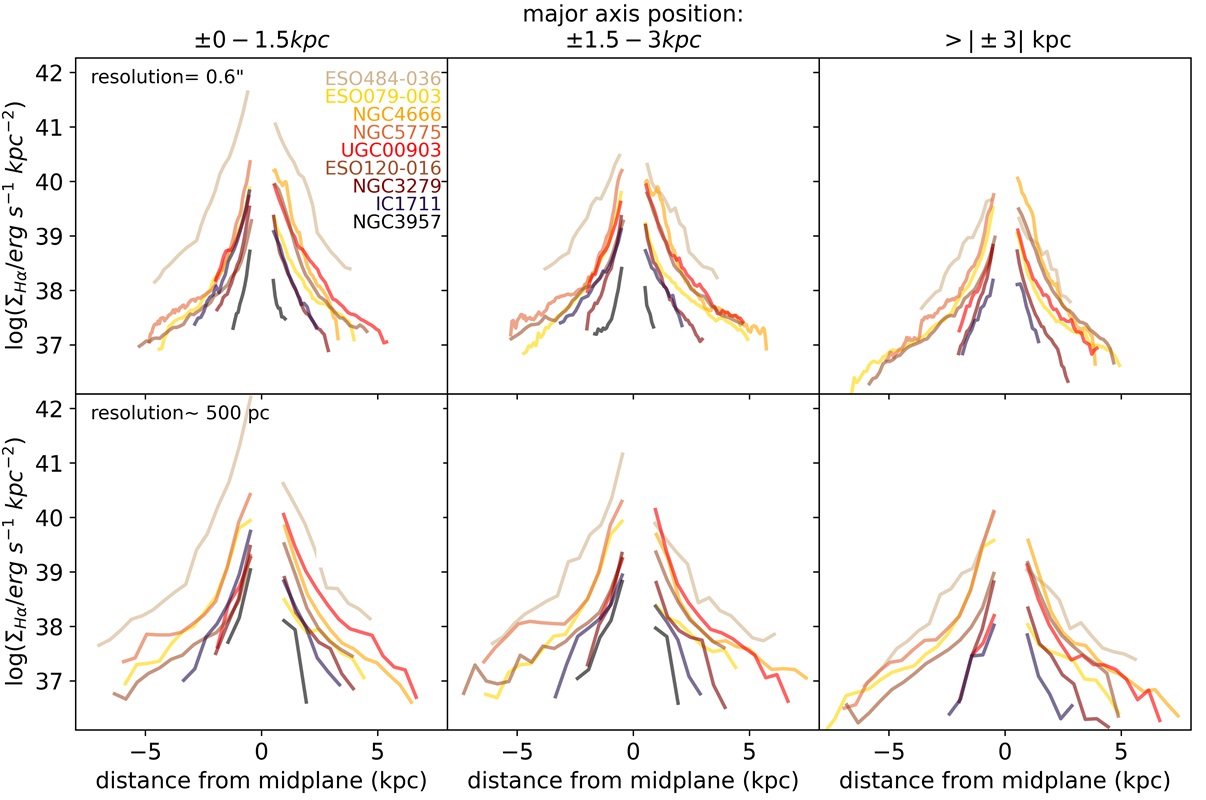}
\caption{Vertical H$\alpha$ surface brightness profiles. Row 1 shows profiles made from 0.6 arcsec resolution data (i.e.: the same resolution shown in Fig.~\ref{fig:maps}), and row 2 shows profiles made from 500~pc resolution data. Each column displays profiles made from the 3 extraplanar regions defined in Section~\ref{sec:region_definition} (see Fig.~\ref{fig:regions}). For NGC~4666 and NGC~5775, only one side of their minor axis profile is displayed, due to the vertically limited field of view for these galaxies. Galaxies in the legend are listed in the order of decreasing SFR. Galaxy centers $\pm500$~pc are omitted to better emphasize the extraplanar gas. 
}
\label{fig:Ha_sb_profiles}
\end{figure*}

Figure~\ref{fig:Ha_sb_profiles} shows vertical profiles of H$\alpha$ surface brightness, $\Sigma_{\rm H\alpha}$, for each galaxy. We make profiles using 0.6~arcsec resolution data (top row) and 500~pc resolution data (bottom row).  Profiles are generated with the average of $\Sigma_{\rm H\alpha}$ in each row, within each subregion, and restricting the averages to only include $S/N>5$ on the emission line in individual spaxels. In all profiles the central $\pm$500~pc, the disk region discussed above, is omitted. The profiles, therefore, concentrate on extraplanar emission only.

In Fig.~\ref{fig:Ha_sb_profiles} there is a trend that highest SFR galaxies have high surface brightness $\Sigma_{\rm H\alpha}$ profiles, and lowest SFR galaxies have lowest surface brightness profiles. This is most clear for the left and center panels, which represent emission closer to the galaxy center (on the major-axis). In the left-hand panels, galaxies ESO~484-036, NGC~4666, NGC~5775, ESO~079-003, ESO~120-016, and UGC~00903 exhibit emission extending up to $\sim$5~kpc from their midplanes. In contrast, the lowest SFR galaxies (NGC~3957, NGC~3279, and IC~1711) only reach $\sim$2-3~kpc above the midplane. Assuming that H$\alpha$ flux is a mass tracer, and that gas generating extraplanar emission originates in the galaxy there are many reasons why we would expect the low SFR systems to have less extended profiles of H$\alpha$ emission. 

It is interesting to note that IC~1711 has a low $\Sigma_{H\alpha}$ values, that do not extend as far as the high SFR galaxies. We have discussed how the kinematics and emission line ratios of IC~1711 were more similar to high-SFR galaxies, yet this similarity does not extend to the H$\alpha$ brightness. If one assumes that the biconical shape to the $\sigma_{H\alpha}$ on this galaxy suggests an outflow, then the H$\alpha$ surface brightness would suggest a very low mass in the outflow of IC~1711. 


\section{Emission line ratio diagnostics of ionisation mechanisms}
\label{sec:BPT line diagnostics}

The Baldwin–Phillips–Terlevich (BPT) diagram \citep{Baldwin1981} is a widely used tool to investigate the ionization mechanism driving emission line fluxes \citep[reviewed in][]{Kewley.2019}. IFU studies have further applied BPTs to prove spatial variation in ionisation across various star-forming galaxies \citep[e.g.][]{Ho2014,Davies2014,Sharp&BlandHawthorn2010,S´anchez.2020}. 

In this section, we create BPT diagrams to study the ionisation properties of extraplanar gas in our sample. We will use these diagrams to compare galaxies and galaxy regions (defined in Section~\ref{sec:region_definition}) in our sample. We will also use these diagrams to check differences between galaxies that are higher SFR galaxies than from lower. In resolved studies of face-on galaxies, \cite{Reichardt2025} finds that outflows are more common for specific star formation rate, $SFR/M_{star}>0.1$~Gyr$^{-1}$. For a galaxy with, M$_{\star}\sim5\times10^{10}$~M$_{\odot}$ (median GECKOS stellar mass), $SFR/M_{star}\sim 0.1$~Gyr$^{-1}$ corresponds to SFR$\sim5$~M$_{\odot}$~yr$^{-1}$.

In this study, we construct two types of emission line ratio diagrams for each galaxy: [NII]/H$\alpha$ against [OIII]/H$\beta$;  and [SII]/H$\alpha$ against $\sigma_{\mathrm{H}\alpha}$. As a test, we created BPT diagrams of [SII]/H$\alpha$ against [OIII]/H$\beta$ to identify AGN activity signatures in our line-ratrio data. Though these extra BPT diagrams are not shown in this paper, we did not find clear AGN signatures using [SII] in place of [NII]. In each diagram, we only include pixels with $S/N>5$ detections for [NII], H$\alpha$, [OIII], and H$\beta$. 

Traditionally, BPT diagrams use theoretical classification curves  \citep[e.g.][]{Kewley2001,Kauffmann2003} that define regions of the parameter space dominated by different ionisation mechanisms. However, the shape and position of these curves depend on the hardness of the local ionizing radiation field \citep{Kewley2019}. We must consider this effect for our galaxy sample, as we expect harder radiation fields to occur naturally in lower attenuated extraplanar regions. To account for varying ionizing field strengths, we use the variable classification curve by equation~5 of \cite{kewley2013}  in our [NII]/H$\alpha$ against [OIII]/H$\beta$ BPT diagrams. We, therefore, plot two curves, that with smaller line ratios represents the standard star forming sequence from \cite{kewley2013}; the solid curve with larger line ratios represents a harder field. 

Fig.~\ref{fig:BPT_Nii_group1} shows 2D contours of the [NII]/H$\alpha$-[OIII]/H$\beta$ diagram for our targets. The contours represent kernel density estimates of the distribution of pixels from corresponding galaxies. We show levels that include 66$\%$ and $99\%$ of the probability mass for each contour set. We separate the galaxies into the following three subsets: SFR$\geq 5$~M$_{\odot}$~yr$^{-1}$ galaxies (blue contours), SFR$<5$~M$_{\odot}$~yr$^{-1}$ galaxies (red contours) and IC~1711 (green contours). We have separated IC~1711 due to the differences in morphology of both $\sigma_{H\alpha}$ and ionization maps described in the previous sections. Each subplot displays a shock model grid (black dashed line) and two thermal ionization models (solid lines). The shock models taken from Tables 6 and 8 of the MAPPINGSIII simulation paper by \cite{Allen2008}. Thermal emission models are generated using equation~5 in \cite{kewley2013}.  The region between these shocks and thermal emission is often described as the ``mixing sequence'', where the relative contributions of thermal and non-thermal ionization are ambiguous. We show [NII]/H$\alpha$-[OIII]/H$\beta$ diagrams for all individual targets in the Appendix. Figures in the appendix show individual spaxels, which are colored by distance to the galaxy midplane.

\subsection{Line ratios for extraplanar gas of high SFR galaxies}
\label{sec:BPTs for high SF galaxies}

\begin{figure*}
\includegraphics[width=\textwidth]{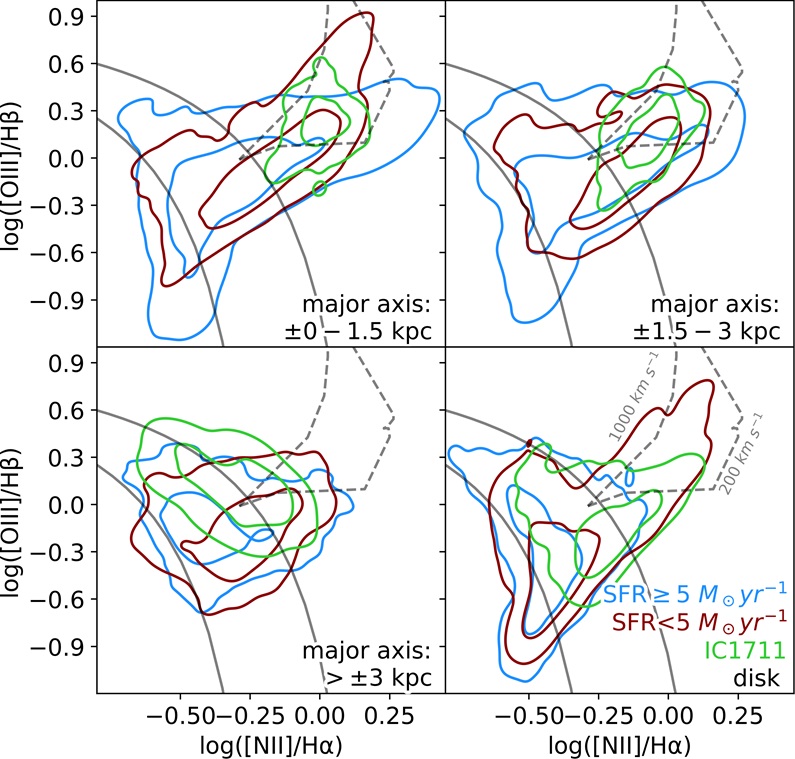}
\caption{A set of [NII]/H$\alpha$ against [OIII]/H$\beta$ BPT 2D histograms, combining all pixels with $S/N>7$ in each line from our sample. We divide the sample into galaxies with SFR$\geq$5~M$_{\odot}$~yr$^{-1}$ galaxies (blue contours), IC~1711 (green contours) and those with SFR below this (red contours). Contours represent the top 66\% and 99\% of pixels in each category. Separate BPT diagrams are shown for the 4 galaxy regions defined in Section~\ref{sec:region_definition} (see Fig.~\ref{fig:regions}).  The spacing of the two solid lines reflect an increasing radiation-field, with the line for larger line ratios representing the harder field \citep{kewley2013}. The black dashed line represents a shock model grid from MAPPINGSIII simulations \citep{Allen2008}, covering shock velocities between 200~km~s$^{-1}$ and 1000~km~s$^{-1}$, which represent the boundaries of available models.}
\label{fig:BPT_Nii_group1}
\end{figure*}


\begin{figure*}
\includegraphics[width=\textwidth]{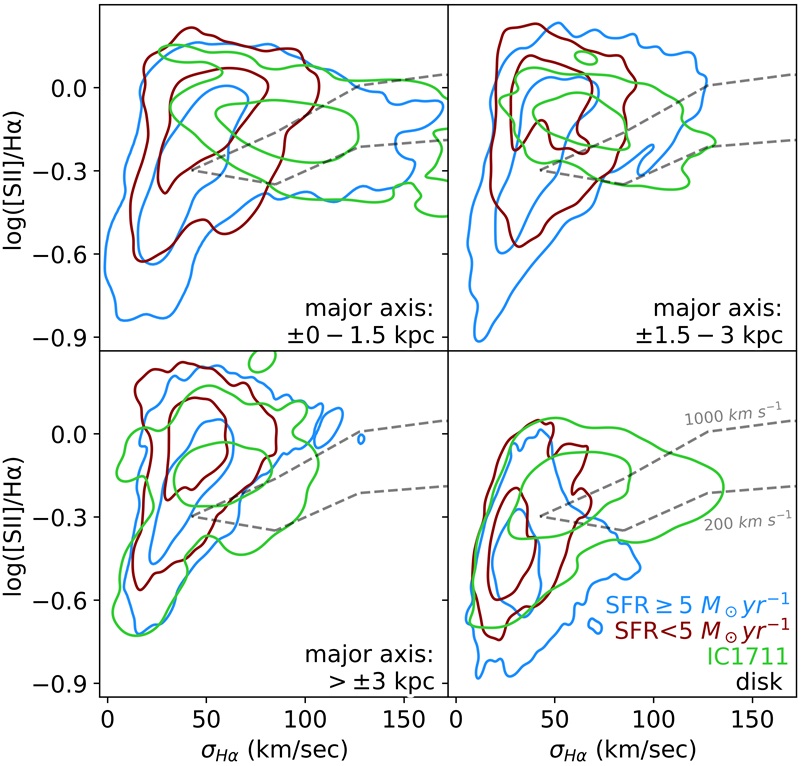}
\caption{Similar galaxy groups as shown in Fig.~\ref{fig:BPT_Nii_group1}, here plotting $\sigma_{\mathrm{H}\alpha}$ against [SII]/H$\alpha$. These diagrams follow the same general structure as Fig.~\ref{fig:BPT_Nii_group1}.}
\label{fig:BPT_FWHM_group1}
\end{figure*}

We now examine emission line diagnostics for galaxies with SFR$\geq 5$~M$_{\odot}$~yr$^{-1}$, indicated as the blue lines in Fig.~\ref{fig:BPT_Nii_group1} and Fig.~\ref{fig:BPT_FWHM_group1}. This group consists of ESO~484-036, NGC~4666, NGC~5775, and ESO~079-003, each shown independently in Appendix Figs.~A1, A2, A3 and A4. The first three of these targets have known outflows \citep{Veilleux_2005,Dahlem1997,Heald2006}. ESO~079-003 exhibits a pronounced X-shape structure in [NII]/H$\alpha$ (Fig.~\ref{fig:maps}), a high velocity dispersion cone structure (Fig.~\ref{fig:maps} and Fig.~\ref{fig:maps_500pc}), and very extended H$\alpha$ emission (Fig.~\ref{fig:Ha_sb_profiles}). ESO~079-003 is therefore consistent with several indicators of having a strong galactic wind. Therefore, we interpret this group as a sample of outflow-dominated systems.

The blue contours in the top two panels of Fig.~\ref{fig:BPT_Nii_group1} both show emission line ratios for the SFR$\geq 5$~M$_{\odot}$~yr$^{-1}$ galaxies that extend beyond the thermally excited regime and into the non-thermal regime,  consistent with shocks ($v_{\rm shock}\sim 200-300$~km~s$^{-1}$). This includes both the region designed to cover the central part of the outflow ($\pm$0-1.5~kpc) and the region that covers the edge of the outflow ($\pm$1.5-3.0~kpc). The plots in the Appendix show that there is a correlation of larger values of emission line ratios (especially [NII]/H$\alpha$) with distance to from the galaxy midplane. We note that the majority of spaxels fall outside the region occupied by the \cite{Allen2008} shock models.

For the positions more distant from the galaxy center (major-axis position $> 3$~kpc) the majority of spaxels are consistent with thermal emission. There is only a small spur of gas that extends along the mixing sequence, albeit with lower [OIII]/H$\beta$ than observed in the central regions of the extraplanar gas. These results are consistent with differences in ionisation near to (above) the galaxy center, with decrease strength in ionisation for gas more distant to the center of the galaxy.

Fig.~\ref{fig:BPT_FWHM_group1} compares [SII]/H$\alpha$ to the H$\alpha$ line velocity dispersion for our SFR$\geq 5$~M$_{\odot}$~yr$^{-1}$ galaxies (blue contours). We do not include ESO~484-036 in this figure, due to a skyline overlaping with [SII]. 
The highest velocity dispersons are found in the regions above the galaxy centers ($\pm$0-1.5~kpc), reaching $\sigma_{\rm H\alpha}\sim$150~km~s$^{-1}$.  In all extraplanar regions, $\sigma_{\rm H\alpha}$ positively correlates with [SII]/H$\alpha$ for $\sigma_{\rm H\alpha}\lesssim60-70$~km~s$^{-1}$. This corresponds to [SII]/H$\alpha\sim1$. For larger velocity dispersions ($\sigma_{\rm H\alpha}\sim80-150$~km~s$^{-1}$) there is not significant increase in [SII]/H$\alpha$.

\cite{Kewley.2019} argue that shocks are probable drivers of emission line if (1) shock-sensitive line ratios, such as [SII]/H$\alpha$, correlate with dispersion, and (2) the velocity dispersion is larger than $\sim$80~km~s$^{-1}$. We find a mixture of evidence for shocks in gas above the plane in the GECKOS high SFR targets. We, likewise, point out that only a small fraction of spaxels show both high $\sigma_{\rm H\alpha}$ and a positive correlation between $\sigma_{\rm H\alpha}$ and [SII]/H$\alpha$. Additionally, Fig.~\ref{fig:BPT_Nii_group1} shows that the most elevated line ratios remain near to the lower end of the \cite{Allen2008} shock model grid. This may be due to the models being designed for ISM conditions, rather than outflows. Alternatively, shock fronts likely occur on much smaller scales than our spatial resolution (50-200~pc). Recent work finds that clouds are typically on the order of $10$~pc in width \citep{Fisher2024}. If shocks occur on these scales, and are embedded in a non-shocked gas, then our observations would blend the two together. This could modify the measured emission line ratio. 

\cite{Chisholm2017} finds negligible contribution of shocks from UV absorption lines in outflows, though it is important to note that absorption lines and emission lines do not trace gas in the same way. Nevertheless, the lack of difference in the nature of the [SII]/H$\alpha$-$\sigma_{\rm H\alpha}$ relationship, and relatively few spaxels that overlap with shock models, may simply reflect a minimal role of shocks in the ionization of wind gas. 

\citet{McPherson2023} discusses the observational effect for observing outflows with relatively low spectral resolution instruments ($R\sim1000-3000$), like MUSE. Observations of nearby outflows with higher spectral resolution observe split emission lines in outflows \citep{Westmoquette2008,2011MNRAS.414.3719W}. The line-splitting is understood to result from the two sides of an expanding cone of gas. If an expanding cone is observed with lower spectral resolution, this would appear as higher velocity dispersion for a single Gaussian component. An interpretation of the shape of the $[SII]/H\alpha-\sigma_{\rm H\alpha}$ relationship could be that at low dispersion there is a physical link between the ionisation and gas kinematics, but at higher velocity dispersion the velocity dispersion reflects the expanding cone, which is unrelated to the ionisation mechanism. 

\subsection{Line ratios for extraplanar gas of main-sequence galaxies} \label{sec:BPTs for low SF galaxies}


In Fig.~\ref{fig:BPT_Nii_group1} we also show the 66\% and 99\% contours of the [NII]/H$\alpha$-[OIII]/H$\beta$ values for galaxies with SFR$<5$~M$_{\odot}$~yr$^{-1}$ (i.e. ESO~120-016, NGC~3279, NGC~3957, UGC~00903), excluding IC~1711. The lower SFR systems are shown as red contours. These galaxies lie within $\pm$0.3~dex of the main-sequence SFR for their given mass, with  lower SFR/M$_*$ than is expected for strong outflows (see Table \ref{tab:GECKOS_sample_table}). We therefore interpret these galaxies as typical main-sequence star-forming galaxies in the local Universe, that are not typically associated to large-scale outflows. 


The first thing we notice is that, in general, the contours for extraplanar gas of the low- and high-SFR occupy similar regions of [NII]/H$\alpha$-[OIII]/H$\beta$ parameter space. The majority of spaxels, indicated by the inner 66 percentile contours, of both subsets of galaxies overlap well. While the distributions are similar, there are subtle differences between the distribution of extraplanar gas in the [NII]/H$\alpha$-[OIII]/H$\beta$  diagram for high and low SFR galaxies. The high SFR systems reach ratios of [NII]/H$\alpha\sim3$, while the main-sequence systems are rarely far above [NII]/H$\alpha\gtrsim1$. Conversely, the extraplanar emission of the main-sequence targets reach higher values of [OIII]/H$\beta$ (upper left panel) and a more substantial fraction of the spaxels fall in the shock model region.  

In all 4 regions of the main-sequence galaxies (red contours), we find emission line ratios that extend from the thermal ionisation region of the BPT diagram, along the mixing sequence and into the region of the \cite{Allen2008} shock models. This is similar to the results of \cite{GonzalezDíaz2024}, who studies the extraplanar emission of 9 edge-on galaxies with SFR similar to the main-sequence targets of our sample. They find emission line ratios consistent with significant contribution from shocks.


The red contours in Fig.~\ref{fig:BPT_FWHM_group1} show velocity dispersion-[SII]/H$\alpha$ distributions from galaxies with SFR$<5$~M$_{\odot}$~yr$^{-1}$. Here the bulk of the low-SFR systems (all except IC~1711) have a very different relationship between [SII]/H$\alpha$ and $\sigma_{\rm H\alpha}$ as compared to the high SFR systems. The low SFR systems cover a similar range of [SII]/H$\alpha$  but do not show the high velocity dispersion tail over the galaxy center that is present in the systems that are known to host strong outflows. This suggests that the differences in increased SFR do not impact the ionisation mechanism heavily, but do in fact lead to differences in the kinematics. This makes sense under the assumption that extraplanar gas is driven by feedback and that higher SFR have a larger mechanical energy in a launch mechanism. This larger energy generates a coherent biconical outflow. Conversely, feedback may drive gas above the plane in the main-sequence systems, but this does not form an expanding bicone.

The elevated values of [NII]/H$\alpha$ and [OIII]/H$\beta$ values in the disk region subplot of Fig.~\ref{fig:BPT_Nii_group1} correspond to NGC~3957 (see the Fig.~A7). This galaxy is different from the rest of the sample in many ways. Firstly, it has the least bright extraplanar gas. Secondly, the extraplanar gas that is present is located in an off-center plume that is counter-rotating with respect to the disk, and has elevated dispersion (see 2nd and 3rd column in Fig.~\ref{fig:maps3} respectively). This counter-rotating plume is not typical for outflows, as outflowing material has often been shown to align closely with the galactic minor axis, and to co-rotate with the galactic disk \citep[e.g. ][]{Shopbell1998,Heald2006}.  An accretion or merger of a smaller galaxy could be consistent with counter-rotating, elevated gas. The merging gas does not need to have the same line ratios of the host, which could lead to this difference. Alternatively, Fig.~\ref{fig:maps} shows that the highest [NII]/H$\alpha$ values in NGC~3957 are associated to the galaxy center, possibly reflecting a contribution from AGN.

\subsection{Emission line ratios for extraplanar gas in IC~1711}

We have previously discussed the unique case, within our sample, of IC~1711. This galaxy has a low SFR, $\sim1$~M$_{\odot}$~yr$^{-1}$, but the extraplanar gas shows a high dispersion cone-shape, which is more commonly found at high SFR. The H$\alpha$ surface brightness of the extraplanar gas is, however, not high and is more appropriate for the low-SFR.  Fig.~\ref{fig:BPT_Nii_group1} and Fig.~\ref{fig:BPT_FWHM_group1} show observations of IC~1711 as green contours. 

Above the galaxy center (major-axis 0-1.5~kpc) and in the adjacent region (1.5-3~kpc) the [NII]/H$\alpha$-[OIII]/H$\beta$ values are centered on the \cite{Allen2008} shock models (see green contours in Fig.~\ref{fig:BPT_Nii_group1}).  Similarly, these regions have high $\sigma_{\rm H\alpha}$ and exclusively higher [SII]/H$\alpha$. The only other galaxy with exclusively high line ratios above the disc plane is ESO484-36, which is the strongest SFR target in the sample and hosts a very strong biconical outflow. ESO484-36 has the brightest H$\alpha$ surface brightness above the midplane. While IC~1711 has relatively weak extraplanar emission, the gas that is present above the plane has very similar ionisation and dynamical properties as the galaxies we understand to be stronger outflows. The emission line ratios are likewise similar to strong outflows. These reason for these similarities, in absence of strong H$\alpha$ flux above the midplane, remain unclear.

\subsection{Galaxy Averages for Extraplanar Emission}

\begin{figure*}
\includegraphics[width=\textwidth]{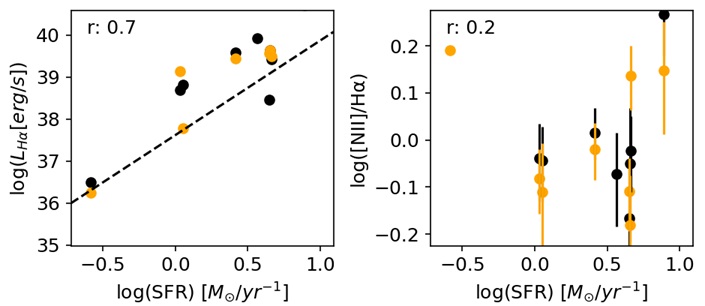}
\caption{Measurements of SFR against extraplanar H$\alpha$ luminosity ($\mathrm{L}_{\mathrm{H}\alpha}$) and SFR against [NII]/H$\alpha$ for the extraplanar regions across our galaxy sample. The extraplanar gas measurements are made for gas at larger distance from the galaxy midplane than $\pm$2~kpc, and only in the central region of the extraplanar emission ($\pm$0-1.5~kpc). This represents the brightest part of the extraplanar gas in all targets, except for NGC~3957 (discussed above). We display measurements made above (black points) and below (orange points) the midplane of each galaxy.  A spearman rank correlation coefficient is indicated in the top left corner of each plot. }
\label{fig:SFR_scatterplots}
\end{figure*}

In Fig.~\ref{fig:SFR_scatterplots} we show the dependence of integrated extraplanar emission ($\mathrm{L}_{\mathrm{H}\alpha}$) and median extraplanar [NII]/H$\alpha$ on the total SFR of each galaxy. The extraplanar gas is measured only at z-axis heights that are greater than $\pm$2~kpc from the galaxy midplane. Also, we restrict the measurements to the central region of the extraplanar gas, which means the inner $\pm$1.5~kpc on the major-axis. We tested use of a wider major-axis range, but it does not significantly change Fig.~\ref{fig:SFR_scatterplots}. We find a strong correlation of galaxy SFR with total extraplanar H$\alpha$ luminosity, with Spearman correlation coefficient of 0.7 and a fitted power law of $\log(L_{H\alpha})=2.3\ \log(\rm SFR)+37.6$. This result is similar to \cite{Lu2023}, who found more extended H$\alpha$ emission in higher SFR galaxies. 

Several authors conclude that the extraplanar H$\alpha$ emission around star forming galaxies is consistent with being driven by feedback \citep[e.g.][]{Levy2019,Lu2023,GonzalezDíaz2024}. Energy injected from supernovae, cosmic rays, and radiation pressure all increase with SFR \citep[reviewed in][]{Thompson.&.Heckman.(2024)}, which would therefore generate brighter gas emission above higher SFR systems. Our observed correlation between H$\alpha$ luminosity and SFR is consistent with this. For supernova-driven winds the energy injection into the outflow scales linearly with the SFR \citep[e.g.][]{Thompson.&.Heckman.(2024)}. If we make a somewhat simplistic assumption that outflow energy scales primarily with mass in the outflow, then we, likewise, expect a linear relationship of outflow mass with SFR \citep[similar to results in][]{Reichardt2025}. The mass of ionised gas depends on $L_{\rm H\alpha}/n_e$, where $n_e$ is the electron density. If $n_e$ of the outflow increases with SFR of the galaxy, then a power-law that is near to $\sim$2 for the relationship between $SFR$ and $L_{\rm H\alpha}$ is needed for mass in the outflow to scale linearly with SFR. This would be consistent with the power-law in Fig.~\ref{fig:SFR_scatterplots}, which implies that our power-law is consistent with extrplanar emission, in these galaxies, being driven by stellar feedback. 

Recently, \cite{MazzilliCiraulo2025} reports the first well resolved $n_e$ profile that extends deep into an outflow ($z\sim6$~kpc), using GECKOS observations of NGC~4666 (included in our sample).  At the least, these works show that high electron density is found in winds, which would be consistent with a linear relationship of outflow mass with SFR in GECKOS targets.

The relationship between [NII]/H$\alpha$ and SFR does not show a simple power-law behavior. Only two galaxies with strong evidence for outflows, ESO~079-003 and ESO~484-036, have elevated extraplanar [NII]/H$\alpha$ line ratios. NGC~3957, which we have previously discussed as having several differences, also has an elevated extraplanar [NII]/H$\alpha$ value. The majority of the sample, however, has low [NII]/H$\alpha$ values that are below unity. The differences in the extremes of the distribution of the emission line diagnostics found in Fig.~\ref{fig:BPT_Nii_group1} and Fig.~\ref{fig:BPT_FWHM_group1}, do not represent the bulk of the gas. This can be seen in the individual plots available in the Appendix, which color points based on distance to the galaxy. These show that the highest line ratios, where these differences emerge, occur in a fainter gas found at a larger distance to the galaxy. 

Fig.~\ref{fig:SFR_scatterplots} also suggests that changes in H$\alpha$ flux do not necessarily drive changes in [NII]/H$\alpha$. While the total flux of H$\alpha$ above the plain appears explainable, the line ratios of extraplanar gas are not so easily described. This may simply reflect that no one phenomenon cleanly explains all ionization mechanism in extraplanar gas around galaxies.

\section{Discussion \& Conclusions}
\label{sec:Discussion}

In this paper, we use observations from the GECKOS survey to analyse the properties of emission lines from ionised gas in galaxies. We use the first nine star forming galaxies observed. We compared extraplanar gas properties of above the midplane of starburst galaxies to those of main-sequence. We find that extraplanar emission is ubiquitous in our sample and often extends well beyond $\pm2$~kpc from the disk midplane. We find that the surface brightness and luminosity of extraplanar gas in galaxies correlates strongly with SFR of the galaxy, consistent with previous studies on the extraplanar emission in galaxies \cite[e.g.][]{Rossa2003,Levy2019,Lu2023}. 

\subsection{Velocity dispersion maps are good indicators of large biconical winds}
Our sample includes 3 galaxies that have previously been identified as hosting outflows (ESO~484-036, NGC~5775 and NGC~4666). Additionally, ESO~079-003 shows extended extraplanar emission with a centralised high velocity dispersion structure and an X-shaped [NII]/H$\alpha$ morphology (shown in Fig.~\ref{fig:maps}). These features are all characteristic of an outflow. These outflow galaxies have $SFR/M_{star}>0.1$~Gyr$^{-1}$,  which previous samples show is more probable to have outflows \citep{Reichardt2025}.

\cite{McPherson2023} argue that elevated velocity dispersion in extraplanar gas is a good indicator of outflows. They also suggest that the conical shape in velocity dispersion maps from integral field spectroscopy observations can distinguish emission associated with an outflow from ambient extraplanar gas. In our 500~pc resolution velocity dispersion maps (Fig.~\ref{fig:maps_500pc}), we find velocity dispersions exceeding 150~km~s$^{-1}$ for our SFR<5~$M_{\odot}~\rm yr^{-1}$ outflowing galaxies. Similar dispersions are observed in other edge-on outflowing galaxies, such as Mrk148 \citep{McPherson2023}, ESO338-04 \citep{Bik2018}, and NGC\,4383 \citep{Watts2024}. Moreover, when the outflow is fully covered in the MUSE mosaic we find a similar biconical shape. By contrast, high dispersions are rare in main-sequence galaxies.

Our results suggest that ionized gas velocity dispersion is an effective diagnostic for identifying candidate outflows for edge-on galaxies with IFS observations. In this case, elevated velocity dispersion with a biconical shape, when seen above a strong starburst (or AGN) in the disk, can trace outflow gas. Inside the conical outflows, the measured ionzed-gas velocity dispersion is expected to be a product of both internal bulk motion of gas and the expansion of the bicone. A sufficiently high spectral resolution ($R\gtrsim7,000$) would resolve the multipeaked emission line profiles needed to characterise this expansion \citep[e.g.][]{Westmoquette2008,McPherson2023}.

\subsection{Ionization of extraplanar gas does not vary heavily across a large range of SFR}

Many studies have explored the ionisation of extraplanar gas in various galaxy samples \citep[e.g.][]{Collins2001,Sharp&BlandHawthorn2010,Ho.et.al..2016,LopezCoba2019,Boettcher2019}. These studies generally find increased ionization at larger distances from the galaxy midplane. This trend
may reflect the growing contribution of shocks as an ionisation mechanism away from the midplane \citep[e.g.][]{Collins2001}, or it could reflect radiation field hardening at larger minor axis heights due to reduced shielding \citep[see discussion in][]{Kewley.2019}. A mixture of both of these effects is also plausible. 

Due to the larger energy and momentum associated to biconical outflows, one might expect that the ionisation properties in strong galactic winds above starburst galaxies to differ from that of the extraplanar gas above main-sequence galaxies. X-shaped [NII]/H$\alpha$ morphologies have long been an ionization signature associated with outflows \citep[review][]{Veilleux_2005}. We see this feature most clearly in ESO~079-003 for our sample (see Fig.~\ref{fig:maps}). In our sample, the largest [NII]/H$\alpha$ and [SII]/H$\alpha$ values are found in the centralized extraplanar gas above ESO~484-036 and ESO~079-003. Both of these galaxies have strong velocity dispersion cones, which may suggest a stronger presence of shocks. Overall, we see the strongest shock-ionization evidence for two of our super main-sequence systems.

Conversely, NGC~4666 and NGC~5775, both outflow galaxies, only sparsely populate the shock region of the BPT parameter space. As seen in Fig.~\ref{fig:SFR_scatterplots}, only 2 of our SFR$\geq$5~M$_{\odot}$~yr$^{-1}$ galaxies have elevated [NII]/H$\alpha$ in the outflow. We find a similar outcome for [SII]/H$\alpha$. Moreover, \cite{Bik2018} find low extraplanar [SII]/H$\alpha$ and low [NII]/H$\alpha$ in outflowing starburst galaxy ESO338-IG04. Despite these ratios, they do identify shocks with the [OI] line at large distances from the galaxy. We will consider GECKOS observations of [OI] in a future work.

We also find that our main sequence ($SFR<5$~$M_{\odot}~\rm yr^{-1}$) galaxies show extraplanar line ratios high enough ([NII]/H$\alpha>$1) to be consistent with shock models by \cite[e.g.][]{Allen2008} (see Figs.~\ref{fig:BPT_Nii_group1}~\&~\ref{fig:BPT_FWHM_group1}). Above the galaxy center the majority of spaxels for high and low SFR are in similar regions of the [NII]/H$\alpha$-[OIII]/H$\beta$ sequence, and main-seqience galaxies span a similar range of [SII]/H$\alpha$ and the that of starbursts. Although some minor differences in line ratio distributions are observeable between the high and low SFR galaxies in Fig.~\ref{fig:BPT_Nii_group1}, our analysis has not revealed any physical significance in these differences. Our findings echo results from \cite{Ho.et.al..2016}, who find that emission-line ratios like [NII]/H$\alpha$ are not sufficient alone for separating galaxies with and without strong winds. The key difference appears to be in kinematics rather than ionisation, which Fig.~\ref{fig:BPT_FWHM_group1} illustrates. 


\cite{Levy2019} argue, based on the kinematics of edge-on galaxies, that feedback drives the majority of extraplanar emission in star forming galaxies, even in main-sequence galaxies. The correlation of extraplanar H$\alpha$ luminosity (Fig.~\ref{fig:SFR_scatterplots}), H$\alpha$ surface brightness (Fig.~\ref{fig:Ha_sb_profiles}) and H$\alpha$ scale-height \citep{Li2016} all with SFR are consistent with a picture in which feedback drives extrplanar emission in all star forming galaxies.  Observations suggest that there is a shallow-slope relating $\Sigma_{\rm SFR}$ to the outflowing velocity of gas launched in winds \citep[e.g.][]{Reichardt2025}. Even modest levels of star formation ($\Sigma_{\rm SFR} \sim 0.1$~M$_{\odot}$~yr$^{-1}$~kpc$^{-2}$) are observed to have ionized gas velocities of $\sim$100-200~km~s$^{-1}$ \citep[e.g.][]{Reichardt2025}, which would potentially drive gas upward and could create a shock against the inner CGM.

Aside from the explanation, from a purely observational view, our results suggest that ionised gas line ratios alone are insufficient to identify the presence of outflows. Using elevated [NII]/H$\alpha$ or [SII]/H$\alpha$ to indicate outflows would likely underestimate the number of systems that show outflows. A combined analysis of velocity dispersion and line-ratios would improve results, but resolved imaging that indicates a conical shape in [NII]/H$\alpha$ and/or $\sigma_{\rm H\alpha}$ provides a more complete selection. 

\subsection{IC~1711: A possible relic outflow}

IC~1711 shows high velocity dispersion above the galaxy center and enhanced emission line ratios in the central part of the extraplanar gas, but lacks bright extraplanar emission. IC~1711 has the second lowest SFR in our sample, with SFR$\sim$1.1~M$_{\odot}$~yr$^{-1}$, and low specific star formation rate, SFR/M$_{star}\sim$0.02~Gyr$^{-1}$. It would be very different than previous observations for this this low star formation to drive an outflow \citep[e.g.][]{Forster-Schreiber_et_al._2019,Reichardt2025}. Moreover, there is no evidence of a presently luminous AGN, see BPT in Appendix. Unless there is a high central $\Sigma SFR$ in this galaxy that our $\Sigma SFR$measurements are insensitive to,  The morphology of the velocity dispersion and line-ratio of IC~1711 contrasts with the patterns seen for the high SFR galaxies, such as ESO~484-036. 

If our $SFR$ measurement is not a significant underestimate, one possible explanation for the behavior of IC~1711 could be that some extraplanar emission may trace relic gas from stronger outflows in the past. \cite{Pillepich2021} show, in TNG simulations, that over pressurized, high ionization cones in Milky Way mass systems persist after the feeding of the galactic wind turns off. In these simulations the outflow is driven by an AGN, and after it turns off a high ionization cone remains for 10-100~Myr. \cite{Pillepich2021} argue that these are similar to the large X-ray and $\gamma$-ray bubbles that extend out from the center of the Milky Way, called ``Fermi Bubbles" \citep{Su2010}. Such bubbles could generate changes to both internal gas kinematics and emission line ratios. \cite{GonzalezDíaz2024} studies emission line ratios of 9 edge-on galaxies with similar SFR to IC~1711. They likewise show at least one target has a similar line-ratio morphology. Deep IFS observations of edge-on low-SFR systems may, therefore, reveal evidence of past outflow activity. This opens an opportunity to study the time evolution of outflows. More work studying these systems would be helpful to understand their nature. 

Above, we discuss that the combined use of gas velocity dispersion and [NII]/H$\alpha$ line ratios is a useful tool to identify outflows. Such a selection would include a galaxy like IC~1711. The additional information from the H$\alpha$ flux, would likely conclude that the mass-outflow rate derived for this target would be very low. It would, therefore, not necessarily be problematic to select targets like IC~1711 with stronger outflow systems, as long as authors considered the a mass tracer, like H$\alpha$ flux.

\vskip 12pt
The total GECKOS sample includes 36 galaxies, $\sim$24 of which are star forming systems. This work represents an initial analysis. Whilst clear kinetic signatures of galactic winds appear in this sample, caveats remain. Future work with the complete GECKOS survey will improve statistics enable analysis of mass-loading, energy-loading, and the detailed physics of galactic winds.

\section*{Acknowledgements}
Based on observations made with ESO Telescopes at the
La Silla Paranal Observatory under program ID 110.24AS. We wish to thank the ESO staff, and in particular the staff at Paranal Observatory, for carrying out the GECKOS observations. This paper makes use of services that have been provided by AAO Data Central (datacentral.org.au). This research has made use of the NASA/IPAC Extragalactic Database (NED; https://ned.ipac.caltech.edu/)operated by the Jet Propulsion Laboratory, California Institute of Technology,
under contract with the National Aeronautics and Space Administration. This research has made use of the NASA/IPAC Infrared Science Archive, which is funded by the National Aeronautics and Space Administration and operated by the California Institute of Technology. Part of this research was conducted by the Australian Research Council Centre of Excellence for All Sky Astrophysics in 3 Dimensions (ASTRO 3D), through project number CE170100013. MM acknowledges support from the UK Science and Technology Facilities Council through grant ST/Y002490/1. FP acknowledges support from the Horizon Europe research and innovation programme under the Maria Skłodowska-Curie grant “TraNSLate” No 101108180.

\section*{Data Availability}

The data is available in the ESO archive. The emission line fits may be available by contacting Deanne Fisher (dfisher@swin.edu.au). 



\bibliographystyle{mnras}
\bibliography{example} 





\appendix


\section{Appendix A}

\subsection{BPT line diagnostics for individual galaxies}
\label{sec:BPT line diagnostics for SFR>5 galaxies}

The following section lists our [OIII]/H$\beta$ BPT diagrams made for each of our our individual galaxies. Compared to Fig.~\ref{fig:BPT_Nii_group1} and Fig.~\ref{fig:BPT_FWHM_group1}, these plots display scatter points for individual pixels. The color of each point represents the corresponding pixel's projected minor axis separation from its galaxy's midplane. We note that there is generally a positive correlation between line ratios, dispersion, and minor axis height in these diagrams.


\begin{figure*}
\includegraphics[height=0.42\textheight]{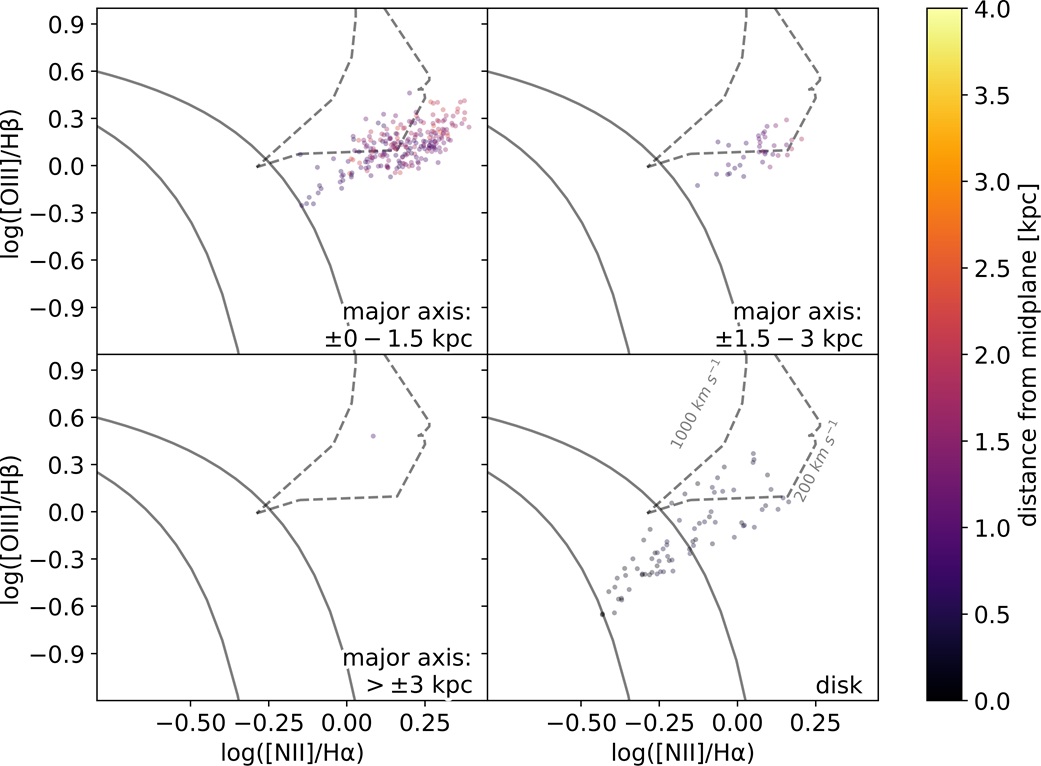}
\caption{A set of [NII]/H$\alpha$ against [OIII]/H$\beta$ BPT diagrams for ESO~484-036. Similar to what is shown in Fig.~\ref{fig:BPT_Nii_group1}, BPT diagrams are given for pixels in the galaxy's major axis position: $\pm$0-1.5~kpc, major axis position: $\pm$1.5-3~kpc, major axis position: >|3|~kpc, and disk regions. Individual pixels are represented by separate scatter points, where the color of each point represents the pixel's minor axis height. Note that the displayed trend lines and grid are the same as the ones shown in Fig.~\ref{fig:BPT_Nii_group1}. This figure is discussed in Section~\ref{sec:BPTs for high SF galaxies}.}
\label{fig:BPT_nii_ESO484-036}
\end{figure*}

\begin{figure*}
\includegraphics[height=0.42\textheight]{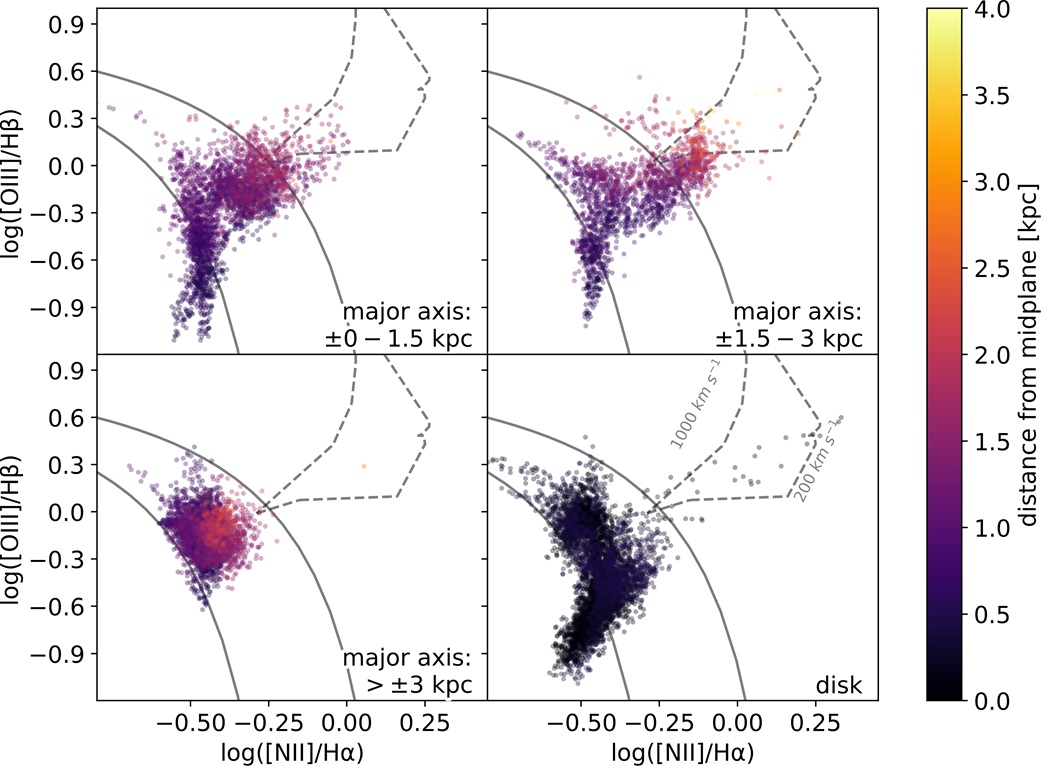}
\caption{A set of [NII]/H$\alpha$ against [OIII]/H$\beta$ BPT diagrams made for NGC~4666. The format is the same as the one used in Fig.~\ref{fig:BPT_nii_ESO484-036}. This figure is discussed in Section~\ref{sec:BPTs for high SF galaxies}.}
\label{fig:BPT_nii_NGC4666}
\end{figure*}

\begin{figure*}
\includegraphics[height=0.43\textheight]{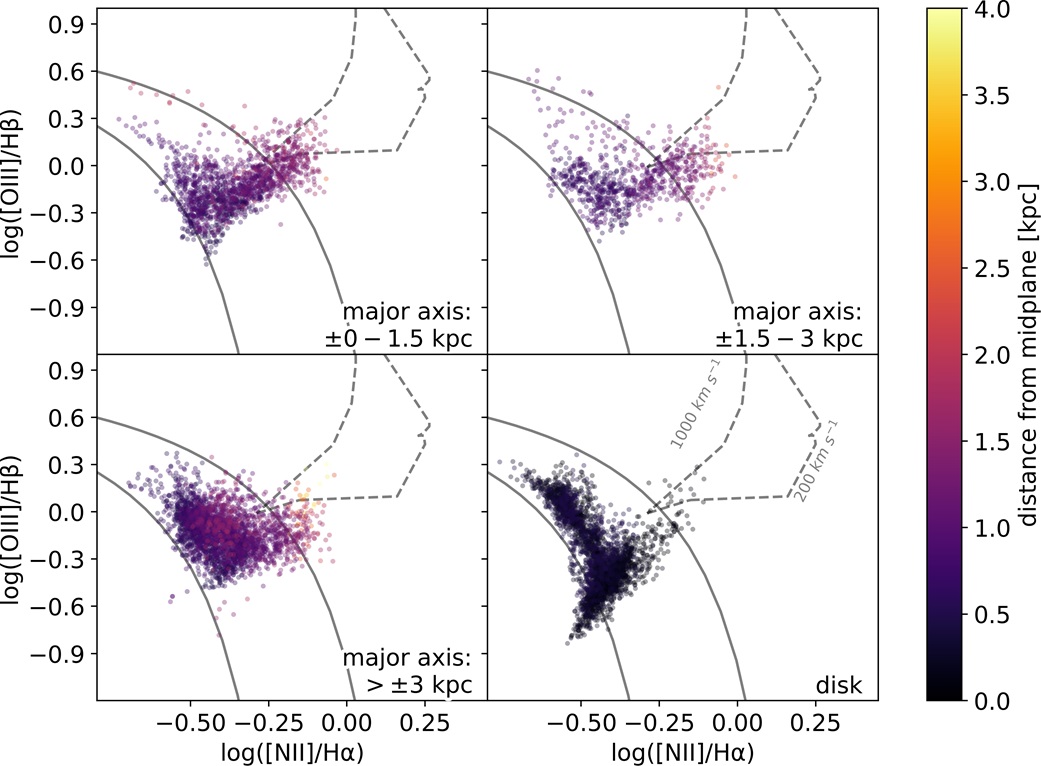}
\caption{A set of [NII]/H$\alpha$ against [OIII]/H$\beta$ BPT diagrams made for NGC~5775. The Format is the same as the one used in Fig.~\ref{fig:BPT_nii_ESO484-036}. This figure is discussed in Section~\ref{sec:BPTs for high SF galaxies}.}
\label{fig:BPT_nii_NGC5775}
\end{figure*}

\begin{figure*}
\includegraphics[height=0.43\textheight]{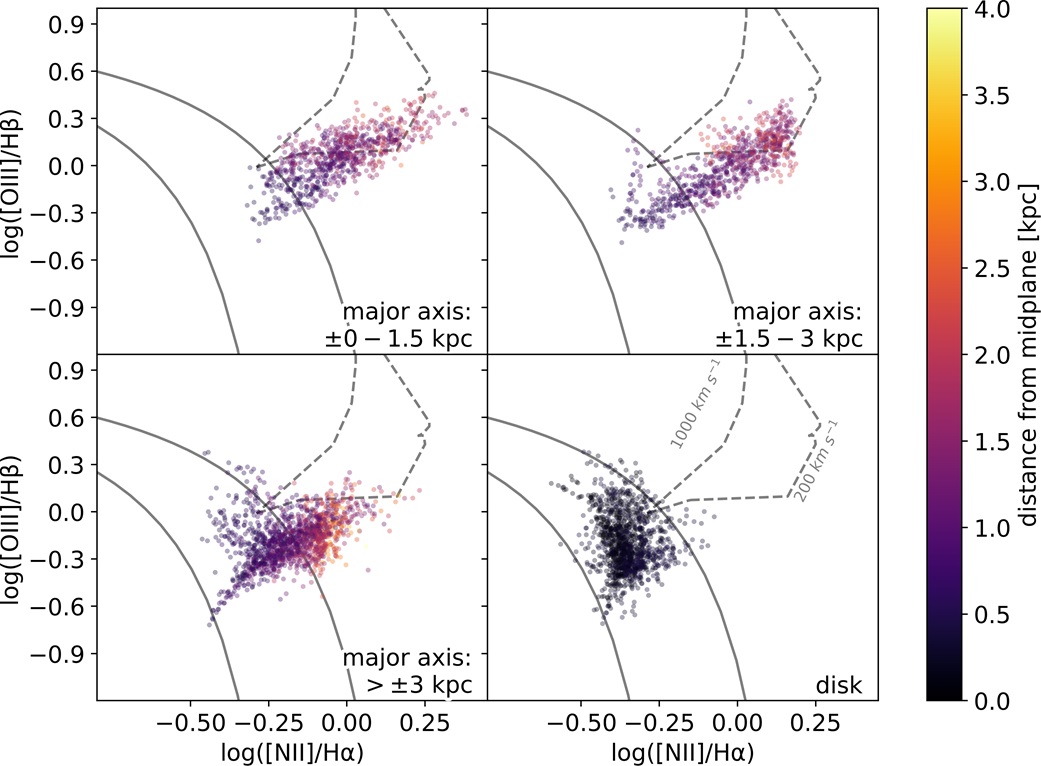}
\caption{A set of [NII]/H$\alpha$ against [OIII]/H$\beta$ BPT diagrams made for ESO~079-003. The format is the same as the one used in Fig.~\ref{fig:BPT_nii_ESO484-036} This figure is discussed in Section~\ref{sec:BPTs for high SF galaxies}.}
\label{fig:BPT_nii_ESO079_003}
\end{figure*}

\begin{figure*}
\includegraphics[height=0.43\textheight]{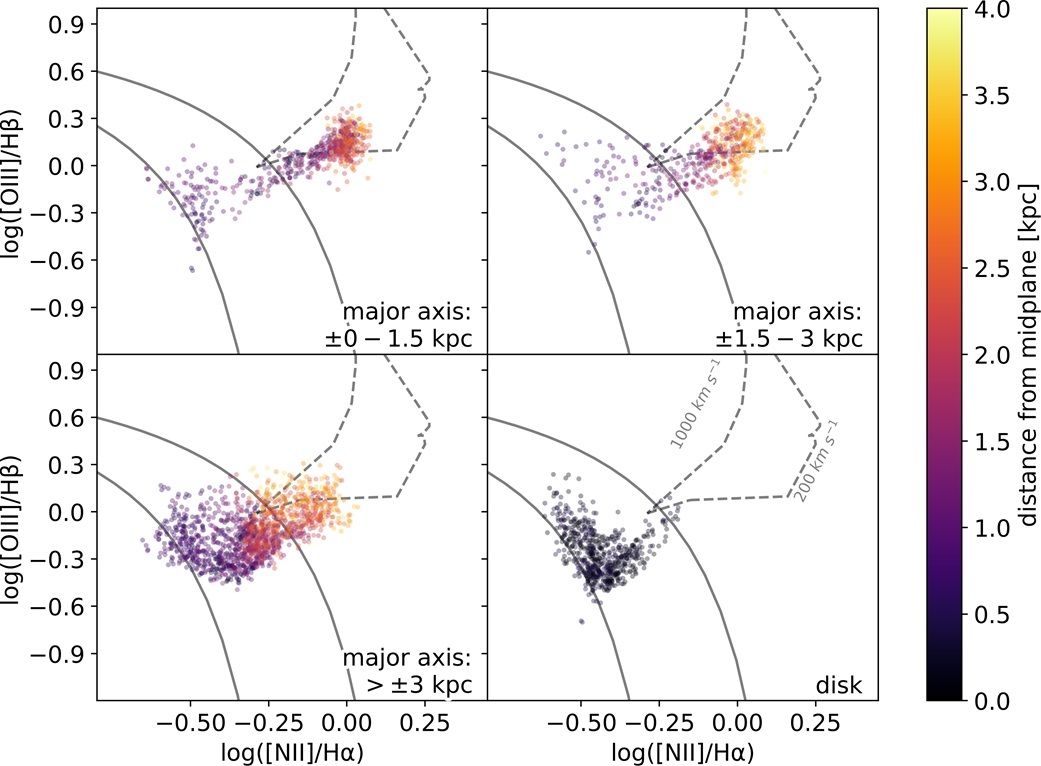}
\caption{A set of [NII]/H$\alpha$ against [OIII]/H$\beta$ BPT diagrams made for ESO~120-016. The format is the same as the one used in Fig.~\ref{fig:BPT_nii_ESO484-036}. This figure is discussed in Section~\ref{sec:BPTs for low SF galaxies}.}
\label{fig:BPT_nii_ESO120_016}
\end{figure*}

\begin{figure*}
\includegraphics[height=0.43\textheight]{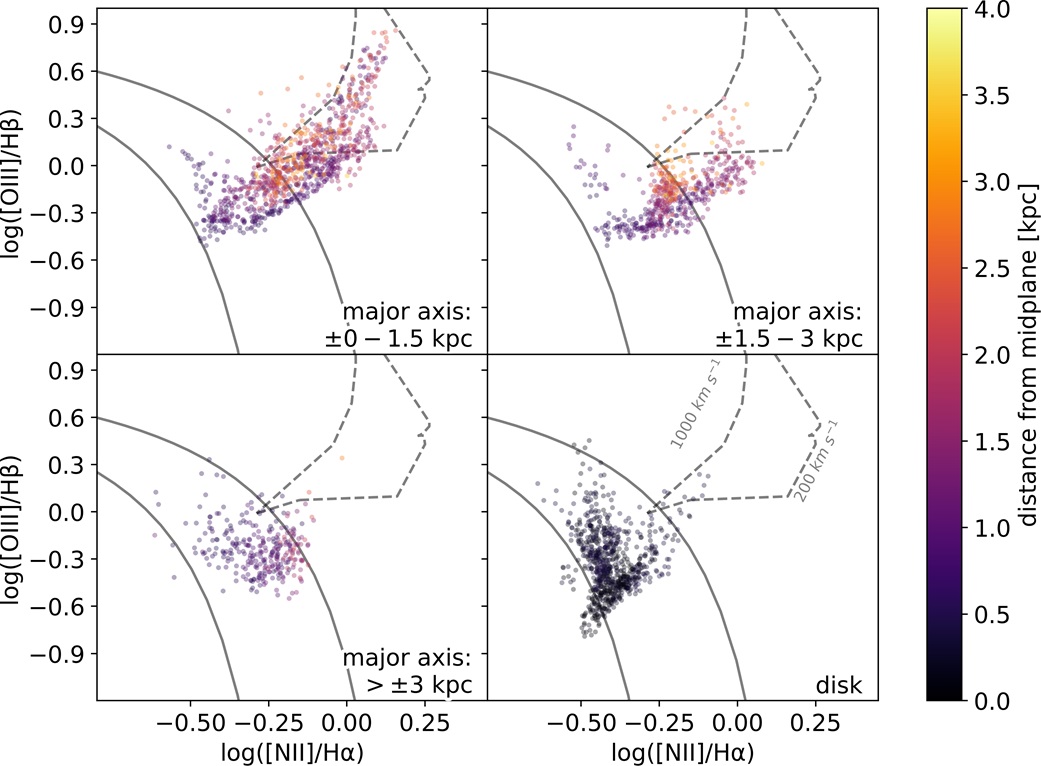}
\caption{A set of [NII]/H$\alpha$ against [OIII]/H$\beta$ BPT diagrams made for UGC~00903. The format is the same as the one used in Fig.~\ref{fig:BPT_nii_ESO484-036}. This figure is discussed in Section~\ref{sec:BPTs for low SF galaxies}.}
\label{fig:BPT_nii_UGC00903}
\end{figure*}

\begin{figure*}
\includegraphics[height=0.43\textheight]{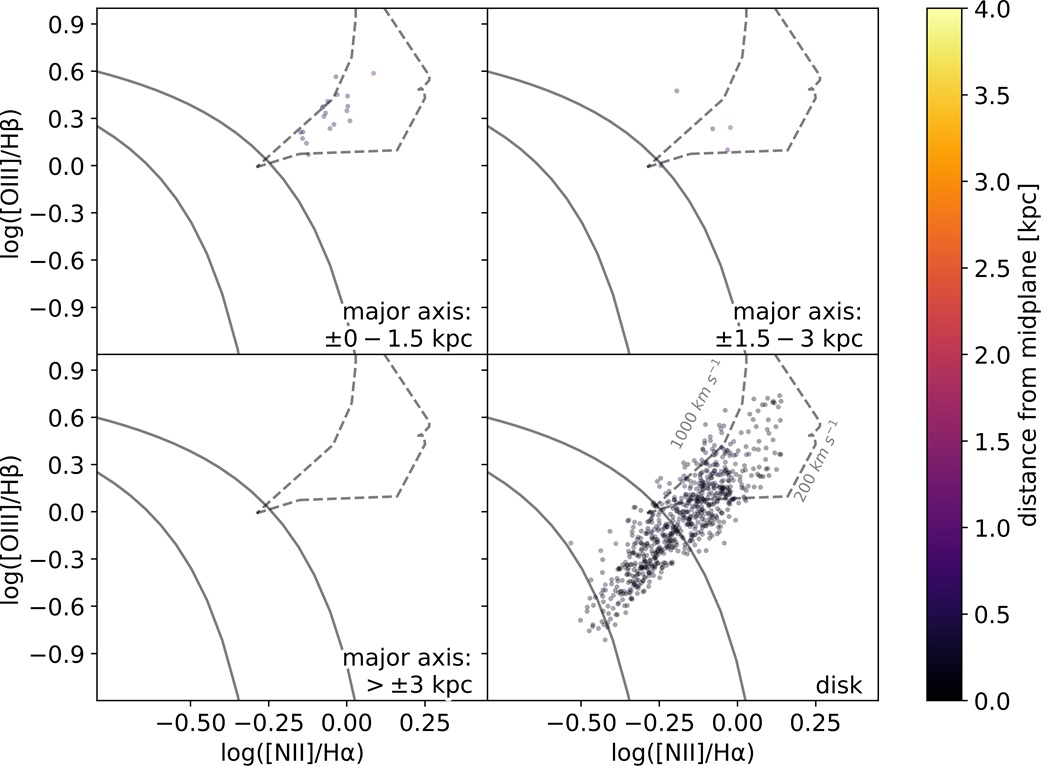}
\caption{A set of [NII]/H$\alpha$ against [OIII]/H$\beta$ BPT diagrams made for NGC~3957. The format is the same as the one used in Fig.~\ref{fig:BPT_nii_ESO484-036}. This figure is discussed in Section~\ref{sec:BPTs for low SF galaxies}.}
\label{fig:BPT_nii_NGC3957}
\end{figure*}

\begin{figure*}
\includegraphics[height=0.43\textheight]{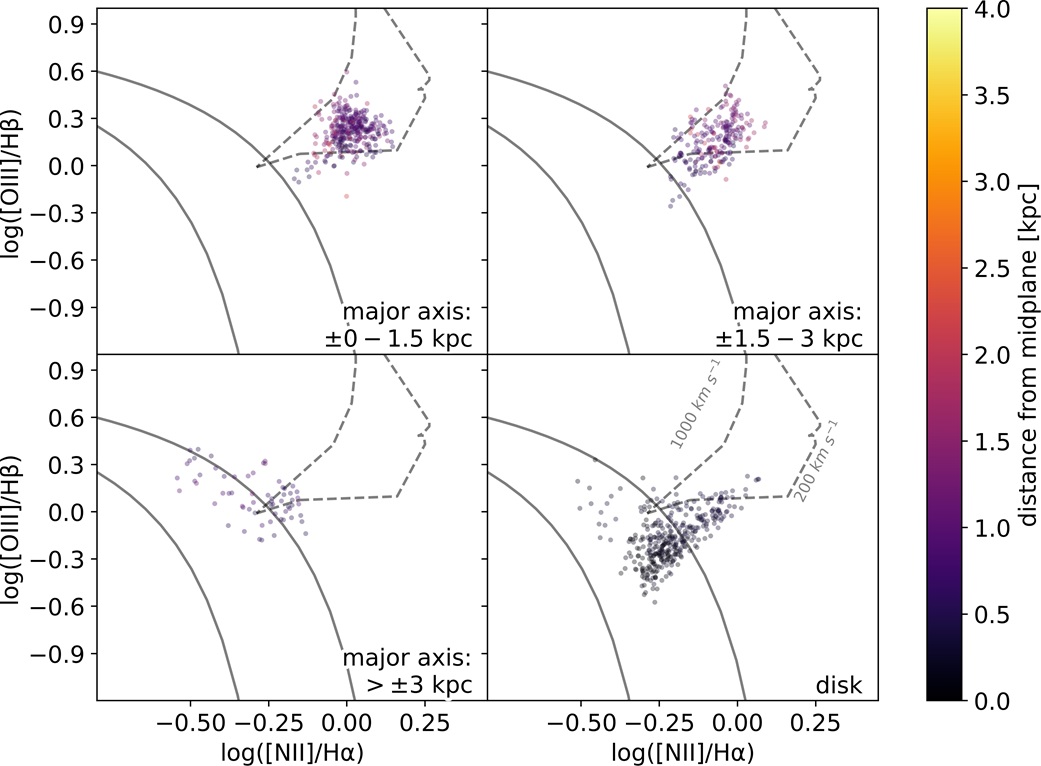}
\caption{A set of [NII]/H$\alpha$ against [OIII]/H$\beta$ BPT diagrams made for IC~1711. The Format is the same as the one used in Fig.~\ref{fig:BPT_nii_ESO484-036}. This figure is discussed in Section~\ref{sec:BPTs for high SF galaxies}.}
\label{fig:BPT_nii_IC1711}
\end{figure*}

\begin{figure*}
\includegraphics[height=0.43\textheight]{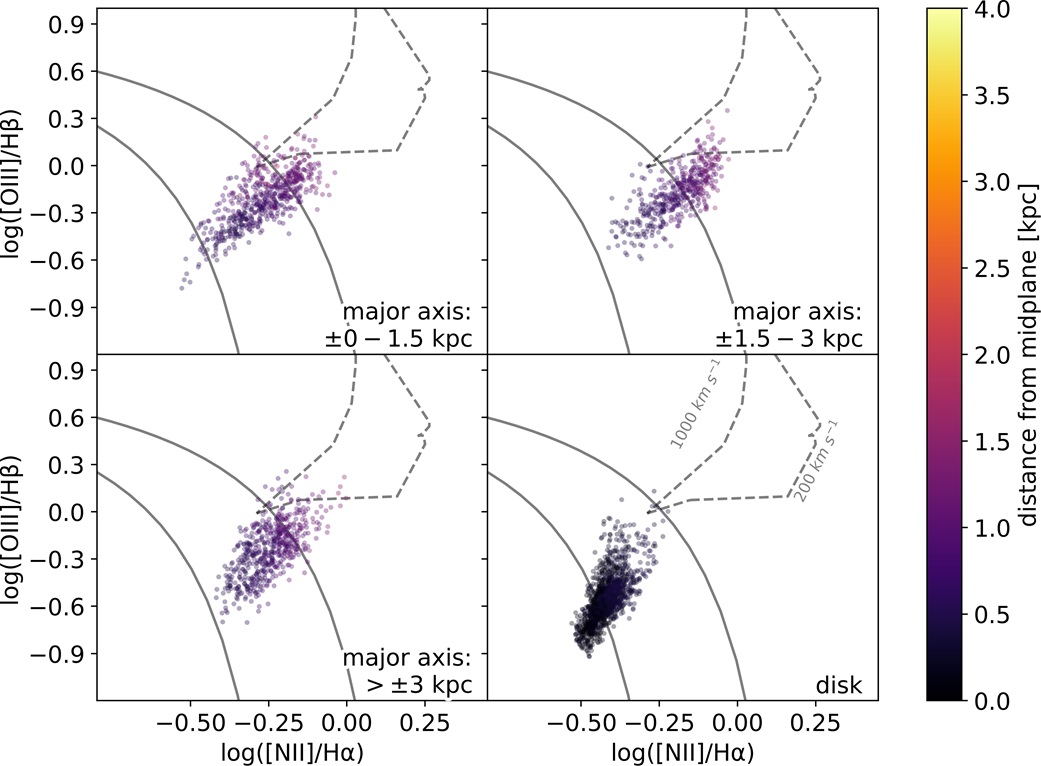}
\caption{A set of [NII]/H$\alpha$ against [OIII]/H$\beta$ BPT diagrams made for NGC~3279. The format is the same as the one used in Fig.~\ref{fig:BPT_nii_ESO484-036}. This figure is discussed in Section~\ref{sec:BPTs for low SF galaxies}.}
\label{fig:BPT_nii_NGC3279}
\end{figure*}

\bsp	
\label{lastpage}
\end{document}